\documentclass[a4paper,12pt]{article}
\usepackage[dvips]{graphicx}
\usepackage{amsmath,amssymb}

\def\gev{{\rm GeV}}
\def\mev{{\rm MeV}}
\def\ev{{\rm eV}}
\def\ten{\textbf{10}}

\def\etal{{\it et al.}}

\newcommand{\beq}{\begin{equation}}
\newcommand{\eeq}{\end{equation}}
\newcommand{\bea}{\begin{eqnarray}}
\newcommand{\eea}{\end{eqnarray}}
\newcommand{\bsub}{\begin{subequations}}
\newcommand{\esub}{\end{subequations} \noindent}
\newcommand{\clean}{\setcounter{equation}{0}}

\def\PRD#1#2#3{Phys. Rev. {\bf D#1} (#2) #3}

\def\NPB#1#2#3{Nucl. Phys. {\bf B#1} (#2) #3}
\def\PTP#1#2#3{Prog. Theor. Phys. {\bf #1} (#2) #3}

\def\PLB#1#2#3{Phys. Lett. {\bf B#1} (#2) #3}
\def\PRL#1#2#3{Phys. Rev. Lett. {\bf #1} (#2) #3}

\begin{document}
\begin{titlepage}
  \begin{flushright}
  \begin{tabular}{l}
    {\bf OCHA-PP-202} \\
    {\bf hep-ph/0302034}
  \end{tabular}
  \end{flushright}
\begin{center}
    \vspace*{1.2cm}
    
    {\Large\bf Neutrino Mass Matrix  \\
\vspace{0.2cm}
          Predicted From Symmetric Texture}\\
\vspace{0.5cm}
    {\large
      Masako {\sc Bando}\footnote{E-mail address:
        bando@aichi-u.ac.jp} 
      and Midori {\sc Obara}\footnote{E-mail address:
        midori@hep.phys.ocha.ac.jp}}   \\
    \vspace{7mm}
    $^1$ {\it Aichi University, Aichi 470-0296, Japan} \\[1mm]
    $^2$ {\it Graduate School of Humanities and Sciences, \\ 
Ochanomizu University, Tokyo 112-8610, Japan}
\end{center}
    \vspace{0.5cm}
\begin{abstract}
\noindent
Within the framework of grand unified theories, 
we make full analysis of symmetric texture to see if such 
texture can reproduce large neutrino mixings, which have recently 
been confirmed by the observed solar and atmospheric neutrino oscillation 
experiments.  
It is found that so-called symmetric texture with 
anomalous $U(1)$ family symmetry with 
Froggatt-Nielsen mechanism does not provide 
a natural explanation of two large mixing angles. 
On the contrary we should adopt "zero texture" 
which have been extensively studied by many authors and 
only this scenario can reproduce two large mixing angles naturally. 
Under such "zero texture"  with minimal 
symmetric Majorana matrix, all the neutrino masses and mixing angles, 
6 quantities, are expressed in terms of up-quark masses, 
$m_t,m_c,m_u$ with two adjustable parameters. 
This provides interesting relations among neutrino mixing angles, 
\bea
\tan^2 2\theta_{12} \simeq \frac{144m_c}{m_t} 
                           \tan^2 2\theta_{23} \cos^2 \theta_{23},
\quad 
\sin^2 \theta_{13}  \simeq  \frac{4m_c}{m_t}\sin^2 \theta_{23}
                            \cos^2 \theta_{12}. 
\nonumber
\eea
Thus $|U_{e3}|$ is predicted to lie within the range $0.01-0.06$. 
Also absolute masses of three neutrinos are predicted almost uniquely. 
This is quite in contrast to the case where bi-large mixings come from 
the charged lepton sector with non-symmetric mass matrix, 
which does not provide the information of neutrino masses.
\end{abstract}
\end{titlepage}
\section{Introduction}
Recent results from KamLAND~\cite{KamLAND} have completely excluded 
all the oscillation solutions for the solar neutrino 
except the Large Mixing Angle solution~\cite{Valle,Fogli,Bahcall,smirnov}. 
This, combined with the present neutrino experiments by 
Super-Kamiokande~\cite{kamioka:2002pe,kamioka:2001nj} and 
SNO~\cite{SNO} have confirmed  
neutrino oscillations with two large mixing angles~\cite{Valle,Smy}:
\begin{eqnarray}
\sin^2 2\theta_{23} > 0.83 ~(99 \%~{\rm C.L.}),    \nonumber \\ 
0.29 \leq \tan^2 \theta_{12} \leq 0.86 ~(99.73 \%~{\rm C.L.}).
\label{expangle}
\end{eqnarray}
The mass squared differences are~\cite{Valle,Nir}
\begin{eqnarray}
\label{expatm}
1.4 \times 10^{-3} < \Delta m^2_{32} <  6.0 \times 10^{-3}~\rm{eV}^2~(99.73 \%~{\rm C.L.}), 
\\ 
5.1 \times 10^{-5} < \Delta m^2_{21} <  9.7 \times 10^{-5}~\rm{eV}^2~(99.73 \%~{\rm C.L.}). 
\label{expsol}
\end{eqnarray}
We are faced with a question; 
 " Why can  such a large difference exist  
 between the quark and lepton sectors? ''
This is  really a challenging question for any particle physicist 
who tries to find grand unified theories (GUTs).  

These neutrino mixing angles are 
expressed in terms of  MNS matrix~\cite{MNS};   
\begin{equation}
V_{MNS}=U_l^{\dagger}U_{\nu},
\label{MNS}
\end{equation}
with $U_l$ and $U_{\nu}$ being the unitary matrices which diagonalize 
the $3\times 3$ charged lepton and neutrino mass matrices, 
$M_l$ and $M_{\nu}$, 
\begin{eqnarray}
U_l^{\dagger} M_l^{\dagger} M_l U_l &=& 
{\rm diag}(m_e^2,m_{\mu}^2,m_{\tau}^2),  \\
U_{\nu}^{\dagger} M_{\nu}^{\dagger} M_{\nu} U_{\nu} &=& 
{\rm diag}(m_{\nu_1}^2,m_{\nu_2}^2,m_{\nu_3}^2), 
\label{mixing}
\end{eqnarray}
respectively.  
Here $M_{\nu}$ is calculated from  
right-handed Majorana neutrino mass matrix, $M_R$, and Dirac neutrino mass 
matrix, $M_{\nu_D}$,
\begin{eqnarray}
M_{\nu}=M_{\nu_D}^T M_R^{-1} M_{\nu_D}. 
\label{seesaw}
\end{eqnarray}
If one want to derive such large mixings without any fine tuning, 
the origin of each of 
the large mixing angle, $\theta_{23}$ and  $\theta_{12}$, 
must be due to either $M_{\nu}$ or $M_l$. 
In the GUT framework larger than $SU(5)$, if it includes Pati-Salam symmetry, 
we have not only a relation between down-type quark and 
charged lepton mass matrices, but also a relation 
between up-type quark and Dirac neutrino mass matrices. 
Thus if one attributes such large mixing angle of Eq.~(\ref{MNS}), 
either up-type or down-type quark mass matrix 
will give important information. 
To be more specific, 
let us adopt susy $SO(10)$ model which has now become more attractive 
since it unifies all the fermions of one family 
together with right-handed neutrino, $\nu_R$. 
Then once we fix each representation of Higgs 
field corresponding to each matrix element,  
$M_l$ and $M_{\nu_D}$ are uniquely determined 
from $M_D$ and $M_U$, respectively. 

In this paper we make semi-empirical analysis on the case in which 
the origin of both those large mixing angles 
comes from the neutrino mass matrix.  We here adopt the so-called 
symmetric four zero texture~\cite{Qfour-zero,Lfour-zero} 
for up-quark mass matrix within 
susy $SO(10)$ GUT framework and examine if they are consistent with 
neutrino experiments with two large mixing angles. 
Here we assume that  each elements of $M_U$ and $M_D$ is dominated by 
the contribution  either from ${\bf 10}$ or ${\bf 126}$, 
the Yukawa coupling of charged lepton are that of the corresponding 
quark multiplied by $1$ or $-3$, respectively. 
More concretely, within symmetric texture the following option 
for $M_D$ has been known to reproduces charged lepton masses 
as well as down quark masses at the same time 
(Georgi-Jarlskog type~\cite{GJ,RRR,DHR,Achiman}); 
\begin{eqnarray}
M_D =
\left(
\begin{array}{@{\,}ccc@{\,}}
 0                 &{\bf 10}     & 0   \\
{\bf 10}           &{\bf 126}    &{\bf 10} \\
 0                 &{\bf 10}     & {\bf 10}
\end{array}\right). 
\label{downmass}
\eea

Now we examine which textures can be cnsistent with 
neutrino two large mixing angles. 
Among various options for configuration 
of ${\bf 10}$ or ${\bf 126}$, 
the following option has been found to be the best type which is  
consistent with the present neutrino experiments, 
\begin{eqnarray}
M_U =
\left(
\begin{array}{@{\,}ccc@{\,}}
 0                 &{\bf 126}           & 0   \\
{\bf 126}           &{\bf 10}            &{\bf 10} \\
 0                 &{\bf 10}            & {\bf 126}
\end{array}
\right).   
\label{upmass}
\eea
This may be compared with Georgi-Jarlskog texture of down-type 
quark mass matrix of the above. 
Further remarkable fact is that once we fix a  proprer option,  
the most economical Majorana neutrino mass matrix with 
two parameters can reproduce all the masses and mixing angles of neutrinos 
consistently with present experiments; even the order 1 coefficients 
are almost uniquely determined from the up-type quark masses only. 
In the next section we make general consideration on symmetric texture. 
We shall see that such mass matrix as derived from anomalous $U(1)$ 
family quantum assignment cannot naturally reproduce large mixing angles. 
According to this fact we here adopt the symmetric four zero texture.   
In section 3 we make numerical calculations and examine various cases. 
We obtain a good candidate for the types of configuration, 
which is investigated in section 4. 
Section 5 is devoted to further discussions.

\section{Symmetric Texture}
\clean
Symmetric texture for fermion mass matrices has been extensively investigated 
by many authors~\cite{DHR,HRR,Nishiura}. 
First let us make an important comment on the Froggatt-Nielsen mechanism~\cite{FN} 
using anomalous $U(1)$ family quantum numbers. 
This is actually an attractive idea for explaining  hierarchy of mass matrices 
observed in quark sectors.  However so far as we take symmetric textures, 
this $U(1)$ cannot naturally reproduce neutrino large mixing angles. 
This might seem to contradict the common understanding that we can 
get any matrix $M_{\nu}$ for any $M_{\nu_D}$ 
by choosing an appropriate $M_R$, namely 
$M_R=M_{\nu_D}M_{\nu}^{-1}M^T_{\nu_D}$. However this is only true if we 
make fine tuning. 

In order to see this, let us take $M_U$ so that each generation of 
left- and right-handed neutrino has anomalous $U(1)$ charge. 
If we restrict ourselves to symmetric texture, namely the family structure 
of left-handed up-type fermions are the same as that of right-handed fermions,  
they have the same $U(1)$ charges, $x_1,x_2, x_3$, respectively. 
Then the Dirac neutrino mass matrix, 
\begin{eqnarray}
S_{\nu_D} \sim \left(
\begin{array}{@{\,}ccc@{\,}}
\lambda^{x_1+x_1}  &   \lambda^{x_1+x_2}  &   \lambda^{x_1+x_3} \\
\lambda^{x_2+x_1}  &  \lambda^{x_2+x_2}   &   \lambda^{x_2+x_3}  \\
\lambda^{x_3+x_1}  &   \lambda^{x_3+x_2}  &   \lambda^{x_3+x_3} 
\end{array}
\right)
\sim \lambda^{x_i}  \lambda^{x_j}
\sim \left(
\begin{array}{@{\,}c@{\,}}
\lambda^{x_1} \\
\lambda^{x_2} \\
\lambda^{x_3} 
\end{array}
\right)
\cdot 
\left(
\begin{array}{@{\,}ccc@{\,}}
\lambda^{x_1}  &   \lambda^{x_2}   &   \lambda^{x_3}
\end{array}
\right).
\label{hienu}
\end{eqnarray}
If we write the inverse of $M_R$  as, 
\begin{eqnarray}
M_R^{-1} \sim \left(
\begin{array}{@{\,}ccc@{\,}}
    D_{11} &  D_{12}     &   D_{13}   \\
    D_{12}  &  D_{22}     &   D_{23}  \\
    D_{13} &   D_{23}   &   D_{33} 
\end{array}
\right), 
\label{arbiR}
\end{eqnarray}
we easily get the following form from Eq. (\ref{seesaw}), 
\begin{eqnarray}
 [ M_{\nu}]_{ij}
\sim 
\lambda^{x_i} \left(\sum\lambda^{x_k}D_{kl}\lambda^{x_l}\right)
\lambda^{x_j}\quad 
\rightarrow \quad 
  M_{\nu}
\sim 
\left(\sum D_{kl}\cdot \lambda^{x_k+x_l}
\right)
\cdot 
S_{\nu_D}.
\label{nuarb}
\end{eqnarray}
This clearly indicates that the resultant matrix $M_{\nu}$ is always 
proportional to the original hierarchical mass matrix. 
Hence it is impossible to get neutrino large mixing  
angles unless the dominant term  of some element vanishes accidentally 
by making fine tuning. 
This is true even if we adjust the order of scales in $M_R$. 
However if some matrix elements of Eq.~(\ref{hienu}) is equal to zero, 
namely if we take texture zero, it can yield large mixing angles 
in the neutrino mass matrix as we shall see below. 

According to the above observation, 
we adopt the following semi-empirical textures for up- and down-type 
quark mass matrices at the GUT scale~\cite{Nishiura} 
\footnote{Here we neglect the CP phases, since they do not change 
largely to the final result. Also we neglect small terms 
using $m_u\ll m_c\ll m_t,\  m_d\ll m_S \ll m_b$.},
\bea
M_D &=& \left(
\begin{array}{@{\,}ccc@{\,}}
0 & \sqrt{\frac{m_d m_s m_b}{m_b-m_d}} & 0 \\ 
\sqrt{\frac{m_d m_s m_b}{m_b-m_d}} & 
m_s & \sqrt{\frac{m_d m_b (m_b-m_s-m_d)}{m_b-m_d}} \\ 
0 & \sqrt{\frac{m_d m_b (m_b-m_s-m_d)}{m_b-m_d}} & m_b-m_d 
\end{array}
\right) \nonumber \\
&\simeq& m_b  \left(
\begin{array}{@{\,}ccc@{\,}}
0 & \frac{\sqrt{m_d m_s}}{m_b} & 0 \\ 
\frac{\sqrt{m_d m_s}}{m_b} &\frac{ m_s}{m_b} &  \sqrt{\frac{m_d}{ m_b}} \\ 
0 & \sqrt{\frac{m_d}{ m_b}} &1
\end{array}
\right),  
\label{Md}
\eea
which reproduces beautifully the down-quark masses and mixing 
as well as charged lepton masses by taking the configuration 
of Eq.~(\ref{downmass}). 
As for the up-quark mass matrix, which we need to get neutrino mass matrix, 
we also take the following form, 
\bea
M_U \simeq m_t \left(
\begin{array}{@{\,}ccc@{\,}}
0 & \frac{\sqrt{m_u m_c}}{m_t} & 0 \\ 
 \frac{\sqrt{m_u m_c}}{m_t} & \frac{m_c}{m_t} &  \sqrt{\frac{m_u}{ m_t}} \\ 
0 & \sqrt{\frac{m_u}{ m_t}} & 1 
\end{array}
\right)
\equiv m_t\left(
\begin{array}{@{\,}ccc@{\,}}
0 & a_u & 0 \\ 
a_u & b_u & c_u \\ 
0 & c_u & 1
\end{array}
\right),
\label{Mu}
\eea
which, together with the down-type texture of Eq. (\ref{Md}), reproduces 
all the observed quark masses as well as CKM mixings.  
However we have not yet found so far 
which configuration of Higgs representations 
should be chosen to give proper neutrino masses and mixings.  
There are 16 textures 
as to which representation is dominated in each component. 
\begin{table}
\caption{Classification of the up-type mass matrices, $M_U$ and $M_{\nu_D}$.}
\begin{center}
\setlength{\tabcolsep}{3pt}\footnotesize
\begin{tabular}[t]{c|c|c|c|c}\hline
Type & Texture & &type & Texture  \\ \hline 
$A_1$ & 
$\left(
\begin{array}{@{\,}ccc@{\,}}
0 & \textbf{126} & 0 \\
\textbf{126} & \textbf{126} & \textbf{126} \\
0 & \textbf{126} & \textbf{126} 
\end{array}
\right)^{\mathstrut}_{\mathstrut}$ &
  &   
$A_2$ & 
$\left(
\begin{array}{@{\,}ccc@{\,}}
0 & \textbf{126} & 0 \\
\textbf{126} & \textbf{126} & \textbf{126} \\
0 & \textbf{126} & \ten 
\end{array}
\right)^{\mathstrut}_{\mathstrut}$  
    \\  \hline
$B_1$ & 
$\left(
\begin{array}{@{\,}ccc@{\,}}
0 & \ten & 0 \\
\ten & \textbf{126} & \textbf{126} \\
0 & \textbf{126} & \textbf{126} 
\end{array}
\right)^{\mathstrut}_{\mathstrut}$ & 
 &
$B_2$ & 
$\left(
\begin{array}{@{\,}ccc@{\,}}
0 & \ten & 0 \\
\ten & \textbf{126} & \textbf{126} \\
0 & \textbf{126} & \ten 
\end{array}
\right)^{\mathstrut}_{\mathstrut}$  
  \\  \hline
$C_1$ & 
$\left(
\begin{array}{@{\,}ccc@{\,}}
0 & \textbf{126} & 0 \\
\textbf{126} & \textbf{10} & \textbf{126} \\
0 & \textbf{126} & \textbf{126} 
\end{array}
\right)^{\mathstrut}_{\mathstrut}$ & 
 & 
$C_4$ & 
$\left(
\begin{array}{@{\,}ccc@{\,}}
0 & \textbf{126} & 0 \\
\textbf{126} & \textbf{10} & \textbf{126} \\
0 & \textbf{126} & \ten 
\end{array}
\right)^{\mathstrut}_{\mathstrut}$  
  \\  \hline
$C_2$ & 
$\left(
\begin{array}{@{\,}ccc@{\,}}
0 & \ten & 0 \\
\ten & \ten & \textbf{126} \\
0 & \textbf{126} & \textbf{126} 
\end{array}
\right)^{\mathstrut}_{\mathstrut}$ & 
  & 
$C_3$ & 
$\left(
\begin{array}{@{\,}ccc@{\,}}
0 & \ten & 0 \\
\ten & \ten & \textbf{126} \\
0 & \textbf{126} & \ten 
\end{array}
\right)^{\mathstrut}_{\mathstrut}$  
  \\  \hline

$F_1$ & 
$\left(
\begin{array}{@{\,}ccc@{\,}}
0 & \textbf{126} & 0 \\
\textbf{126} & \textbf{126}
& \textbf{10} \\
0 & \textbf{10} & \textbf{126} 
\end{array}
\right)^{\mathstrut}_{\mathstrut}$ & 
  &
$F_4$ & 
$\left(
\begin{array}{@{\,}ccc@{\,}}
0 & \textbf{126} & 0 \\
\textbf{126} & \textbf{126} 
& \textbf{10}_2 \\
0 & \textbf{10} & \ten 
\end{array}
\right)^{\mathstrut}_{\mathstrut}$  
  \\  \hline
$F_2$ & 
$\left(
\begin{array}{@{\,}ccc@{\,}}
0 & \ten & 0 \\
\ten & \textbf{126} & \ten \\
0 & \ten & \textbf{126} 
\end{array}
\right)^{\mathstrut}_{\mathstrut}$ & 
&
$F_3$ & 
$\left(
\begin{array}{@{\,}ccc@{\,}}
0 & \ten & 0 \\
\ten & \textbf{126} & \ten \\
0 & \ten & \ten 
\end{array}
\right)^{\mathstrut}_{\mathstrut}$  
  \\  \hline
$S_1$ & 
$\left(
\begin{array}{@{\,}ccc@{\,}}
0 & \textbf{126} & 0 \\
\textbf{126} & \textbf{10} & \textbf{10} \\
0 & \textbf{10} & \textbf{126} 
\end{array}
\right)^{\mathstrut}_{\mathstrut}$ &
 &
$S_2$ & 
$\left(
\begin{array}{@{\,}ccc@{\,}}
0 & \textbf{126} & 0 \\
\textbf{126} & \ten & \textbf{10} \\
0 & \textbf{10} & \ten 
\end{array}
\right)^{\mathstrut}_{\mathstrut}$  
  \\  \hline
$A_3$ & 
$\left(
\begin{array}{@{\,}ccc@{\,}}
0 & \ten & 0 \\
\ten & \ten & \ten \\
0 & \ten & \textbf{126} 
\end{array}
\right)^{\mathstrut}_{\mathstrut}$ & 
 &
$A_4$ & 
$\left(
\begin{array}{@{\,}ccc@{\,}}
0 & \ten & 0 \\
\ten & \ten & \ten \\
0 & \ten & \ten 
\end{array}
\right)^{\mathstrut}_{\mathstrut}$  

  \\  \hline


\end{tabular}
\end{center}
\label{textureclass}
\end{table}%
All possible 16 types are listed in Table~\ref{textureclass}. 
Once we fix their types, the Dirac neutrino mass matrix is automatically 
determined as: 
\bea
M_{\nu_D} 
= \left(
\begin{array}{@{\,}ccc@{\,}}
0 & \ast a_u & 0 \\ 
\ast a_u &\ast b_u & \ast c_u \\ 
0 & \ast c_u & \ast
\end{array}
\right)
\equiv m_t\left(
\begin{array}{@{\,}ccc@{\,}}
0 & a & 0 \\ 
a & b & c \\ 
0 & c & d
\end{array}
\right),
\label{ndiracabc}
\eea
with the Clebsch-Gordan (CG) coefficients $\ast$, $1$ or $-3$ 
according to the corresponding types (see Table~\ref{textureclass}). 
As for the right-handed Majorana mass matrix, to which only 
${\bf 126}$ Higgs field couples, we assume the following 
simplest texture
\footnote{
We shall see later that this texture is accidentally 
consistent with the Higgs representation for the 
Dirac neutrino Yukawa couplings. },
\bea
M_R = 
v_{\bf 126}
\left(
\begin{array}{@{\,}ccc@{\,}}
0 & A_R & 0 \\ 
A_R & 0 & 0 \\ 
0 & 0 & D_R
\end{array}
\right)
\equiv m_R 
\left(
\begin{array}{@{\,}ccc@{\,}}
0 & r  & 0 \\ 
r   & 0 & 0 \\ 
0   & 0 & 1
\end{array}
\right).
\eea
with two parameters, $m_R$ and $r$. 
Now our task is to examine which types in Table~\ref{textureclass} 
can reproduce the neutrino masses and mixing angles consistent 
with the experimental data. 
The neutrino mass matrix is now straightforwardly calculated as, 
\begin{eqnarray}
M_{\nu}=
M_{\nu_D}^T M_{\nu_R}^{-1} M_{\nu_D} 
= \left(
\begin{array}{@{\,}ccc@{\,}}
 0                 &\frac{ a^2}{r}           & 0   \\
\frac{ a^2}{r}   &2\frac{ ab}{r}+ c^2 & c (\frac{a}{r}+1) \\
 0  & c (\frac{a}{r}+1) & d^2
\end{array}
\right) \frac{m_t^2}{m_R} . 
\label{mnu}
\end{eqnarray}
where $a,b,c,d$ are proportional to $a_u,b_u,c_u,1$ with coefficients 
$1$ or $-3$ according to the relevant Higgs multiplets ${\bf 10}$ and 
${\bf 126}$, respectively. 

First we estimate the order of magnitudes
\footnote{ Of course we have to adapt the values of quark masses at GUT scale. 
However the mass ratios are almost independent of the scale, except
for those of top quark.}. 
We know that the order of the parameters in Eq.~(\ref{mnu}) above are 
$a\ll b\sim c\ll 1$.  With this in mind we recognize that the first term 
of 2-3 element of $M_{\nu}$ in Eq.~(\ref{mnu}) should be of order of $d^2$, 
namely $ac/r \sim \mathcal{O}(d^2)$, in order to get large mixing angle $\theta_{23}$. 
This fixes the value of $r$, 
\begin{equation}
r \sim \frac{ac}{d^2} \sim  
*\sqrt{\frac{m_u^2m_c}{m_t^3}} \sim10^{-7},  
\label{eq:}
\end{equation}
which is indeed the ratio of the the right-handed Majorana mass of 
3rd generation to those of the first and second generations. 
At this stage we now come to almost the same situation 
as discussed by Kugo, Yoshioka and one of 
the present author~\cite{Bando:1997ns}. 
If we use the neutrino Dirac masses estimated from 
the charged lepton masses, we would have had almost the same order of 
Majorana masses for the 3rd and 2nd generations. 
On the contrary the top quark mass is very huge compared with that of 
charm quark. This is why we need very tiny value $r$. 
Quite interesting is that the same tiny value $r$ provides 
a large 1-2 mixing angle! 
Furthermore this small value of $r$ is very welcome, as has discussed 
there~\cite{Bando:1997ns}; the right-handed Majorana mass of 
the third generation  must become of order of GUT scale while those of 
the first and second generations are of order $10^{8}$ GeV. 
This is quite favorable for the GUT scenario to reproduce 
the bottom-tau mass ratio. 

Now the problem is whether we can naturally reproduce 
the mixing angle $\theta_{12}$ by the present textures. 
Note that at this stage, once we fix configuration type, 
we have no more arbitrary parameters except for overall scale $M_R$. 
With this $r$, $M_{\nu}$ is approximately written as, 
\begin{eqnarray}
M_{\nu} 
= \left(
\begin{array}{@{\,}ccc@{\,}}
 0                 &\frac{ a^2}{r}           & 0   \\
\frac{ a^2}{r}   &\frac{2ab}{r} & \frac{ac}{r} \\
 0  & \frac{ac}{r} & d^2
\end{array}
\right) \frac{m_t^2}{m_R}
\equiv  
\left(
\begin{array}{@{\,}ccc@{\,}}
 0                 &\beta           & 0   \\
\beta   &\alpha     & h \\
 0  & h  & 1
\end{array}
\right) \frac{d^2m_t^2}{m_R} ,  
\label{apmnu}
\end{eqnarray}
with
\begin{equation}
h=\frac{ac}{rd^2}, \qquad \alpha=\frac{2ab}{rd^2}, 
\qquad \beta=\frac{ a^2}{rd^2}. 
\label{defhalbe}
\end{equation}
Since $\beta \ll \alpha$ and $h\sim \mathcal{O}(1)$, 
we can calculate all the neutrino masses and mixings approximately.  
First let us diagonalize the dominant term with respect to the 2-3 
submatrix of Eq. (\ref{apmnu}).  
The rotation angle, $\theta_{23}$ is written as,    
\begin{eqnarray}
\tan^22\theta_{23}&=& \frac{4h^2}{(1-\alpha)^2},  
\label{rot23} 
\end{eqnarray}
through which  $M_{\nu}$ is now deformed as 
\begin{equation}
\left(
\begin{array}{@{\,}ccc@{\,}}
 0                 &\beta           & 0   \\
\beta   &\alpha     & h \\
 0  & h  & 1
\end{array}\right) \quad
\stackrel{\longrightarrow }{\theta_{23}}
\quad 
\left(
\begin{array}{@{\,}ccc@{\,}}
 0       &\beta \cos\theta_{23}  &\beta \sin\theta_{23}  \\
\beta \cos\theta_{23}  &\lambda_2     &0 \\
\beta \sin\theta_{23}  & 0  &\lambda_3
\end{array}\right)
\label{nu23}
\end{equation}
with 
\begin{eqnarray}
\lambda_3 &=& \frac{\alpha+1+\sqrt{(\alpha-1)^2+4h^2}}{2} 
          \equiv  \lambda_{\nu_3},  \\  
\lambda_2 &=& \frac{\alpha+1-\sqrt{(\alpha-1)^2+4h^2}}{2}.
\label{eigenvaluelam}
\end{eqnarray}
From Eq.~(\ref{rot23}) we notice that $\alpha$ should be very close to 1 
in order to get large mixing angle $\theta_{23}$. 
Since we know the experimental bound for $\theta_{23}$, 
\bea
\sin^2 2\theta_{23} > 0.83, \nonumber
\label{eqatomb}
\eea
we can define the following small value $\varepsilon$ as
\bea
\varepsilon^2 \equiv \frac{(1-\alpha)^2}{4h^2} \le 0.205.
\eea
Let us make a rough estimation 
using the up-quark masses at GUT scale within the error~\cite{koide-fusaoka}, 
\begin{eqnarray}
\alpha &=& \frac{2b}{c}=* \frac{2b_u}{c_u}
        =  * \frac{2m_c}{\sqrt{m_um_t}} \sim * \times (1.0-2.4), \\ 
 \beta &=& \frac{\> a \>}{\> c \>}=* \frac{a_u}{c_u}
        =  * \sqrt{\frac{m_c}{m_t}} \sim *\times (0.03-0.05).   
\label{abfrm}
\end{eqnarray}
Hence in order to be consistent with small value $\varepsilon$, 
the values of $\alpha$ must be close to 1,
so the CG coefficient $*$ in $\alpha$ must be $1$, not $-3$ or $-1/3$.
This requires that 2-2 and 2-3 components of $M_U$ must couple to a 
common Higgs representation.  
With this $\varepsilon$, we can approximately rewrite as follows:
\begin{eqnarray}
\lambda_3 
          &\simeq& 1+h-h\varepsilon,  \\  
\lambda_2 
          &\simeq& 1-h-h\varepsilon.    
\label{eigen23}
\end{eqnarray}

Next step is to rotate with respect to the 1-2 submatrix of Eq. (\ref{nu23}), 
\begin{equation}
\tan^22\theta_{12} = 
\Biggl( \frac{2\beta \cos\theta_{23}}{\lambda_2} \Biggr)^2, 
\label{rot12} 
\end{equation}
with which the neutrino mass matrix becomes 
\begin{equation}
\stackrel{\longrightarrow}{\theta_{12}} 
\quad \left(
\begin{array}{@{\,}ccc@{\,}}
\lambda_{\nu_1} & 0 & \beta\sin\theta_{23}\cos \theta_{12}   \\
0 &\lambda_{\nu_2} & \beta\sin\theta_{23}\sin \theta_{12} \\
\beta\sin\theta_{23}\cos \theta_{12} 
 & \beta\sin\theta_{23}\sin \theta_{12}  & \lambda_{\nu_3} 
\end{array}\right), 
\end{equation}
with eigenvalues,   
\begin{eqnarray}
\lambda_{\nu_2} &=& 
\frac{\lambda_2 + \sqrt{\lambda_2^2+4\beta^2\cos^2\theta_{23}}}{2}, \\ 
\lambda_{\nu_1} &=&
\frac{\lambda_2 -\sqrt{\lambda_2^2+4\beta^2\cos^2\theta_{23}}}{2}.
\label{lambdadasshu}
\end{eqnarray}
Now in order to realize large mixing angle  
$\theta_{12}$ appearing in Eq.~(\ref{rot12}), $\lambda_2$ 
in Eq.~(\ref{eigen23}) must become at least of the same order as $2\beta$. 
This requires again $h\sim 1$. It is not trivial that 
the same $h$ which gives large $\theta_{23}$ 
with the fixed values $\alpha$ and $\beta$ also produces large mixing angle 
$\theta_{12}$ at the same time. 
The larger the ratio $2\beta\cos\theta_{23}/\lambda_2$ is, the larger value 
of $\nu_e$-$\nu_{\mu}$ mixing we get. So it would be desired that $\beta$ 
must be as large as possible, namely the coefficient of $a_u$ in Eq. (\ref{ndiracabc}) 
may be desired to be $-3$
\footnote{ 
This may be more desirable if we take account of the recent KamLAND data 
which suggest very large solar neutrino mixing angle. },
which yields $\beta \sim -0.1$.
In this way we have found the following conditions 
for the desired candidate for the type of Higgs representations. 
\begin{center}
{\bf Condition for Higgs configurations}
\end{center}
\begin{enumerate}
\item  {\it The Higgs representations coupled with 2-3 and 2-2 elements 
of $M_U$ must be  same.}
\item  {\it The Higgs representation coupled with 1-2 elements of $M_U$ 
must be as large as possible.} 
\end{enumerate}
As a result there remain the follwoing two types for 
most desirable candidates, 
\begin{eqnarray}
\left(
\begin{array}{@{\,}ccc@{\,}}
 0                 &{\bf 126}           & 0   \\
{\bf 126}           &{\bf 10}            &{\bf 10} \\
 0                 &{\bf 10}            & {\bf 126}
\end{array}\right), \qquad  
\left(
\begin{array}{@{\,}ccc@{\,}}
 0                 &{\bf 126}           & 0   \\
{\bf 126}           &{\bf 10}            &{\bf 10} \\
 0                 &{\bf 10}            & {\bf 10}
\end{array}\right).  
\label{upmasstypes}
\eea
Finally the neutrino masses are given as, 

\begin{equation}
m_{\nu_3} \sim \lambda_{\nu_3} \frac{d^2m^2_t}{m_R}, \quad
m_{\nu_2} \sim \lambda_{\nu_2} \frac{d^2m^2_t}{m_R}, \quad 
m_{\nu_1} \sim \lambda_{\nu_1} \frac{d^2m^2_t}{m_R}.   
\label{applambdadasshu}
\end{equation}
We should check if the same $h$ can also reproduce 
the two neutrino mass squared differences with one remaining parameter $m_R$. 
In the next section we make numerical calculation and 
see how the above observation 
can be actually confirmed. 
\section{Numerical Calculations}
\clean
In this section we make numerical calculations for all the 
possible types for Higgs configuration.  
The forms of $h,\alpha, \beta$ are written in terms of $m_t, m_c, m_u$ 
with a parameter $r$ in Table~\ref{neutrinomass}.
As our input we take the values of up-type quark masses, 
$m_t,m_c,m_u$ at GUT scale obtained by 
Fusaoka and Koide~\cite{koide-fusaoka};
\bea
\label{koidefusaoka1}
m_u &=& 1.04^{+0.19}_{-0.20}~\mev, \\
m_c &=& 302^{+25}_{-27}~\mev, \\
m_t &=& 129^{+196}_{-40}~\gev.
\label{koidefusaoka2}
\eea
\begin{table}
\caption{The forms of $h$, $\alpha$ and $\beta$ with the value of 
$d$ for each type.}
\begin{center}
\setlength{\tabcolsep}{6pt}\footnotesize
\begin{tabular}[t]{c|c|c|c|c}\hline\hline 
Type 
& d 
& $h$ in $\frac{ac}{r}$ unit
& $\alpha$ in $2m_c/\sqrt{m_um_t}$ unit
& $\beta$ in $\sqrt{m_c/m_t}$ unit \\ \hline \hline
$S_1$ &9    &$ -1/3$ & $1$ & $-3$  \\  \hline
$S_2$ & 1   &$ -3$ & $1$ & $-3$  \\  \hline \hline 
$A_1$ &$1$   &$ 1$ & $1$ & $1$  \\  \hline 
$A_2$ &$1$   &$ 9$ & $1$ & $1$  \\  \hline 
$A_3$ & $9$   &$ 1/9$ & $1$ & $1$  \\  \hline  
$A_4$ &$1$   &$1$ & $1$ & $1$  \\  \hline \hline  
$B_1$ &$9$   &$-1/3$ & $1$ & $-1/3$  \\  \hline  
$B_2$ &$1$   &$ -3$ & $1$ & $-1/3$  \\  \hline  \hline 
$C_1$ &$9$   &$1$ & $-1/3$ & $1$  \\  \hline  
$C_2$ & $9$   &$-1/3$ & $-1/3$ & $-1/3$  \\  \hline 
$C_3$ & $1$   &$-3$ & $-1/3$ & $-1/3$  \\  \hline 
$C_4$ &  $1$   &$9$ & $-1/3$ & $1/9$  \\  \hline \hline 
$F_1$ &  $9$   &$-1/3$ & $-3$ & $-3$  \\  \hline 
$F_2$ & $9$   &$1/9$ & $-3$ & $9$  \\  \hline 
$F_3$ &  $1$   &$1$ & $-3$ & $1$  \\  \hline 
$F_4$ &  $9$   &$-3$ & $-3$ & $-3$  \\  \hline 
\end{tabular}
\end{center}
\label{neutrinomass}
\end{table}%
The present experimental data exists for the mixing angles 
$\theta_{23}$, $\theta_{12}$ , $\Delta m^2_{32}$ 
and $\Delta m^2_{21}$ with upper bound for $|U_{e3}|$. 
Of course, as we mentioned in the previous section, we could 
always reproduce the data $\theta_{23}$ by adjusting 
arbitrary parameter $h$ or equivalently $r$. However 
we already know that $h\sim \mathcal{O}(1)$ so we restrict 
$h$ not too far from 1, keeping the region from $0.3$ to $3.0$. 
Nontrivial is to reproduce large mixing angle $\theta_{12}$ 
since in principle we have no parameter at all, and only the 
freedom is the range of the quark masses at GUT scale of  
Eqs.~(\ref{koidefusaoka1})--(\ref{koidefusaoka2}), 
among which $m_t$ is the most sensitive parameter  
and at least known because of its large Yukawa coupling 
causing ambiguity via its evolution to the GUT scale.  
Within those parameter regions more than half of the types are 
excluded by experiments. Still surprising is that only the CG coefficients 
3 or $-1/3$ does work well. Our model has no more arbitrary parameters 
and everything is fixed without any ambiguity. Thus if the errors of 
experimental data are improved, we can check more strictly whether or 
not our predictions agree with experiments. 
We calculate numerically the above quantities and 
compare them with the experimental data. 
According to Table~\ref{neutrinomass}, 
we classify 16 types into $S$, $A$, $B$, $C$ and $F$ classes.  
\subsection{Class $F$}
This class does not satisfy the condition (i) of the previous section: 
the value of $\alpha$ is 
\bea
\alpha 
\sim -6h\, \frac{m_c}{\sqrt{m_u m_t}} 
\sim (3.0-7.2)\, h, 
\eea
which is far from 1. We calculate the atmospheric neutrino mixing angle, 
$\theta_{23}$. The class $F$ is 
those which does not reproduce large mixing at all, unless we take $h$ 
larger than $3.0$ where we can never reproduce large mixing 
angle $\theta_{12}$, as was mentioned in section 2. 
The calculated results are shown in Fig.~\ref{fig:F1sin2atm} 
from which we confirm that solar neutrino mixing angle remain 
almost two order smaller than experimental bound. 
So we do not have large mixing for solar neutrinos even if 
we could adjust very large  parameter $h$. 
\begin{figure}
\begin{center}
\includegraphics[width=5cm,clip]{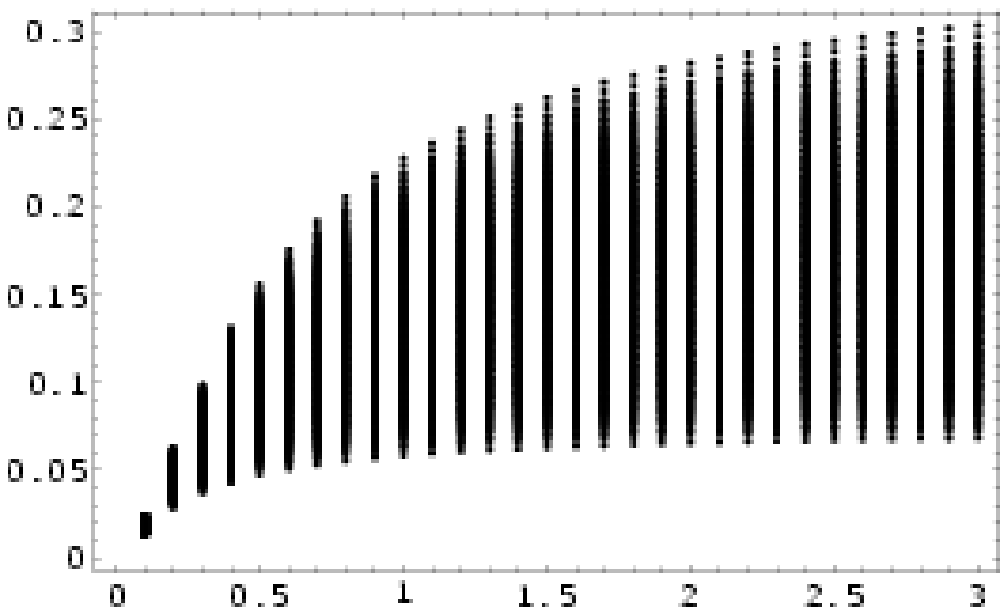}
{\hspace{0.4cm}}
\includegraphics[width=5cm,clip]{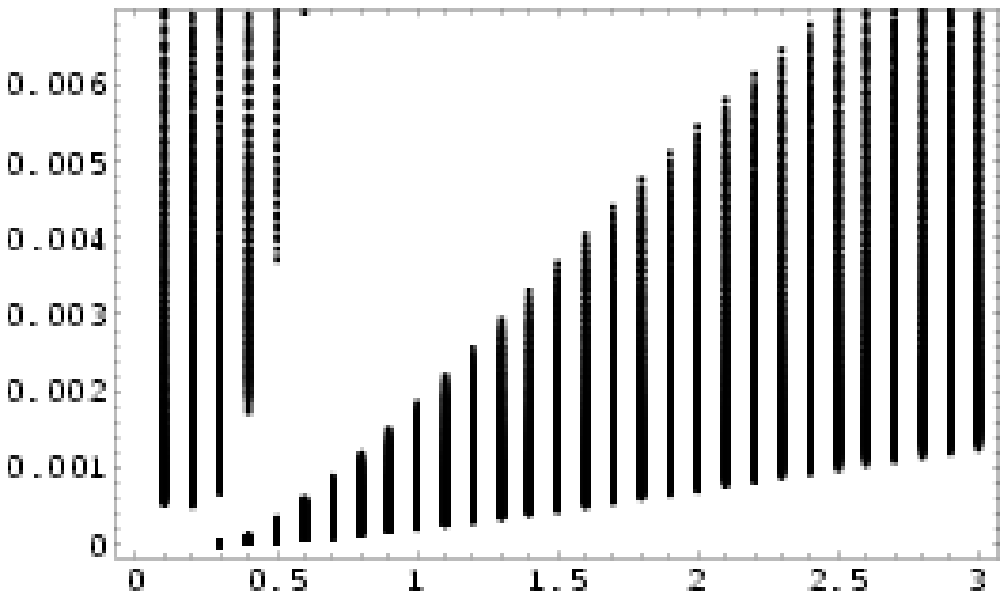}
\end{center}
\caption{Calculated values of $\sin^2 2\theta_{23}$ 
versus $h$ and $\tan^2 \theta_{12}$ versus $h$ in class $F$.} 
\label{fig:F1sin2atm}
\end{figure}%

\subsection{Class $C$}
Next for those which are classified by $C$, the value of $\alpha$ is 
\bea
\alpha 
\sim -\frac{\> 2 \>}{3}h\, \frac{m_c}{\sqrt{m_u m_t}} 
\sim (0.3-0.8)\, h, 
\eea
which is 
not so far from 1.
This range of value makes better
situation than the class $F$. 
We  calculate then the solar neutrino mixing angle, $\theta_{12}$, 
and found that the class $C$ produces too small mixing angle. 
In order that 1-2 mixing of neutrino becomes large, 
$\lambda_2$ and $\beta$ have to be of the same order.
The 2-2 element of $M_{\nu}$ in class $C$,  
namely $\alpha \sim (-2/3) \, b/c$ yields $\lambda_2 \sim 0.1-0.5$.  
However, the value of $\beta$ is 
\bea
\beta 
\sim h\, \sqrt{\frac{m_c}{m_t}} 
\sim (0.03-0.05)\, h, 
\eea
which cannot become the same order as $\lambda_2$. 
Fig.~\ref{fig:C1sin2atm} shows numerical results. 
We see that within the range $h\le 3$, 
the values of $\sin^2 2\theta_{23}$ become larger than 0.83. 
However such $h$ makes the values of $\tan^2 \theta_{12}$ even smaller. 

\begin{figure}
\begin{center}
\includegraphics[width=5cm,clip]{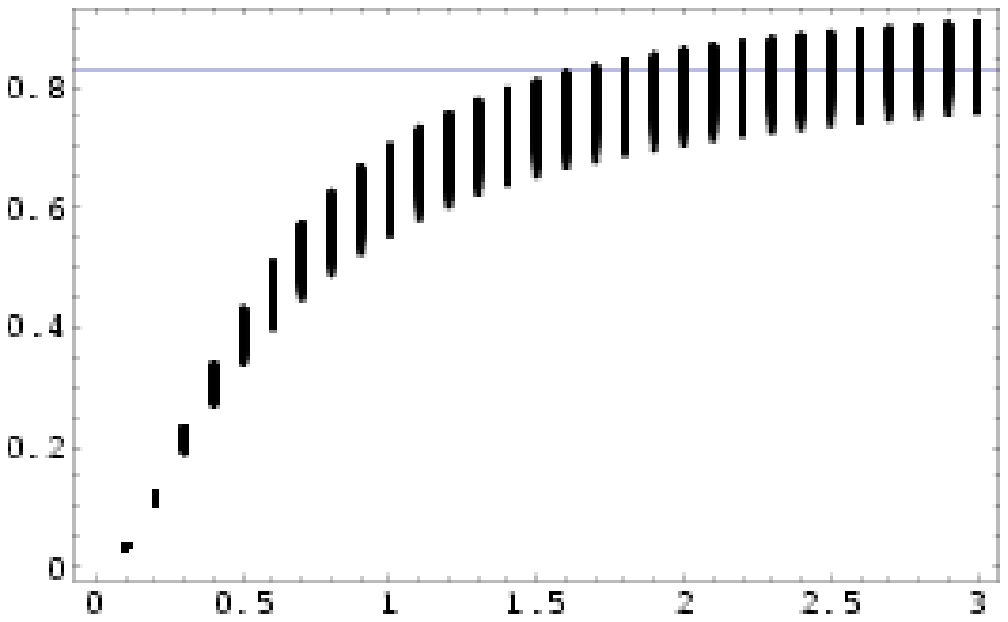}
{\hspace{0.4cm}}
\includegraphics[width=5cm,clip]{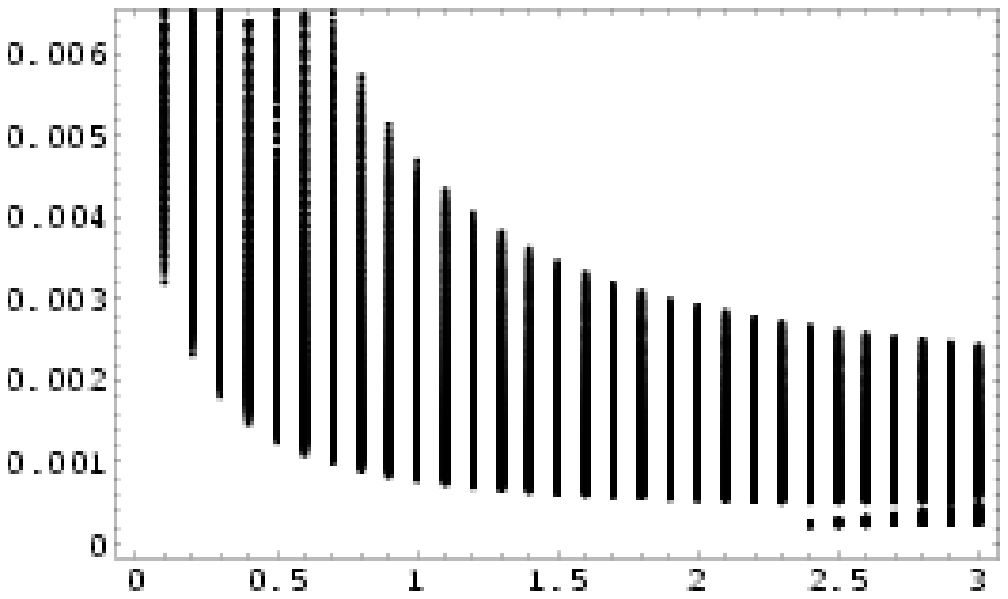}
\end{center}
\caption{Calculated vales of $\sin^2 2\theta_{23}$ 
versus $h$ and $\tan^2 \theta_{12}$ versus $h$ in class $C$.} 
\label{fig:C1sin2atm}
\end{figure}%
\subsection{Class $B$} 

\begin{figure}
\begin{center}
\includegraphics[width=5cm,clip]{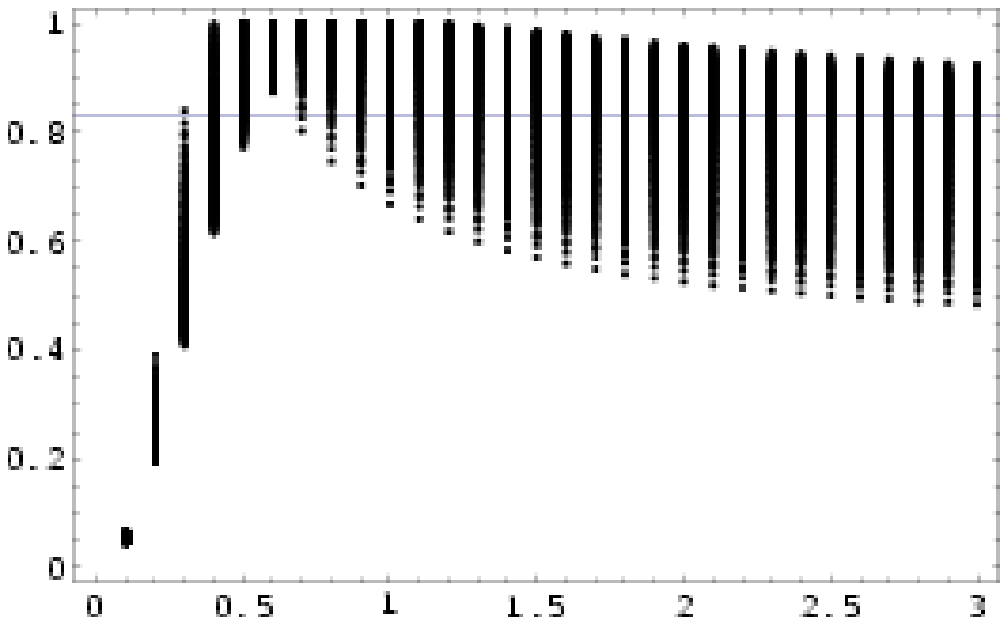}
{\hspace{0.4cm}}
\includegraphics[width=5cm,clip]{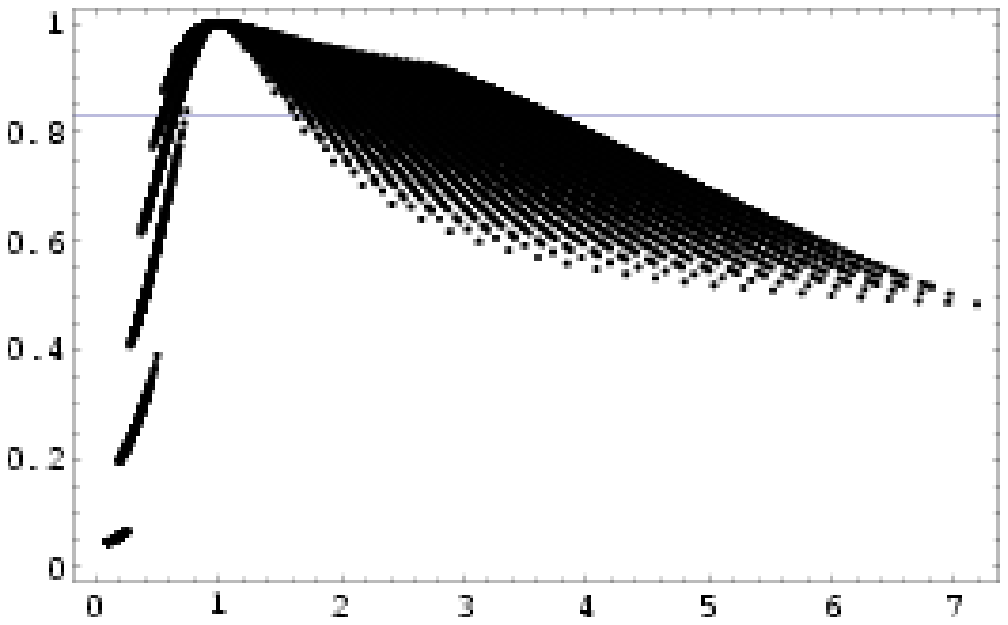}
\end{center}
\caption{Calculated values of $\sin^2 2\theta_{23}$ 
versus $h$ and $\sin^2 2\theta_{23}$ versus $\alpha$ in class $B$.} 
\label{fig:B1sin2atm}
\end{figure}%
This class satisfies the condition (i) and the value of $\alpha$ is
\bea
\alpha 
\sim 2 h\, \frac{m_c}{\sqrt{m_u m_t}} 
\sim (1.0-2.4)\, h.
\eea
Fig.~\ref{fig:B1sin2atm} indicates that 
the calculated results may reach to the bound of 
atmospheric neutrino mixing around $h\sim 1$. 
Since all the parameters are fixed within errors of the up-quark masses 
at GUT scales, we can express $\sin^2 2\theta_{23}$ as a function of 
$\alpha$. The $\alpha$ dependence of $\sin^2 2\theta_{23}$  
is also seen in Fig.~\ref{fig:B1sin2atm}, from which we easily recognize 
that the values of $\sin^2 2\theta_{23}$ becomes larger as $\alpha$ 
approaches closer to 1. 
Thus we can always adjust the parameter $h$ so that it reproduces 
the experimental data of $\sin^2 2\theta_{23}$. 
However once we fix $h$, the type $B$ can hardly reproduce the 
value of mixing angle $\theta_{12}$ 
larger than the experimental bound 0.24.   
\begin{figure}
\begin{center}
\includegraphics[width=5cm,clip]{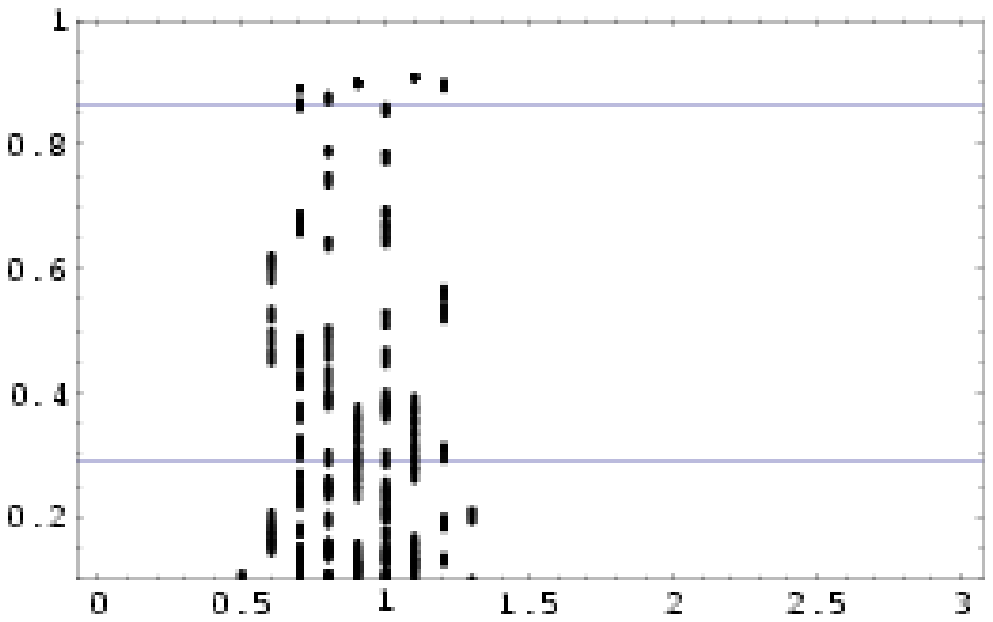}
{\hspace{0.4cm}}
\includegraphics[width=5cm,clip]{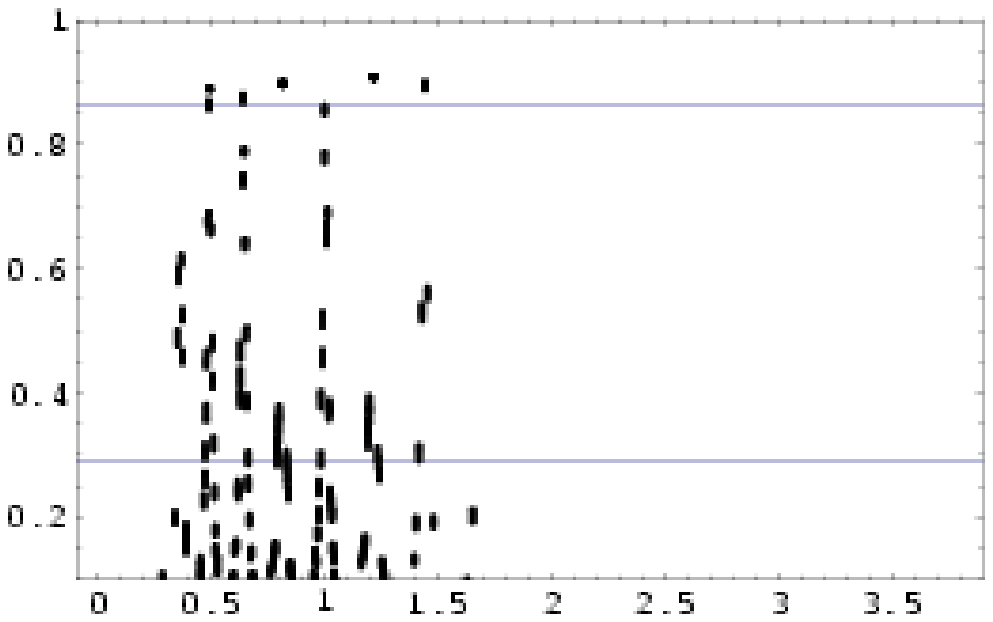}
\end{center}
\caption{Calculated values of $\tan^2 \theta_{12}$ 
versus $h$ and $\tan^2 \theta_{12}$ versus $\alpha$ in class $B$.} 
\label{fig:B1tansol}
\end{figure}%
This can be seen from Fig.~\ref{fig:B1tansol}.
This is because this class does not satisfy the condition (ii)  
and the value of $\beta$ is 
\bea
\beta 
\sim \, \frac{-h}{3}\sqrt{\frac{m_c}{m_t}} 
\sim (0.01-0.017)\, h, 
\eea
which cannot become the same 
order as $\lambda_2$. 
\subsection{Class $A$}
This class satisfies the condition (i) and (ii) and seems to be 
a little easier to reproduce large mixing angle $\theta_{12}$ than class $B$,  
since the value of $\beta$ is 
\bea
\beta 
\sim \, \sqrt{\frac{m_c}{m_t}} 
\sim (0.03-0.05)\, h. 
\eea
Fig.~\ref{fig:A1sin2atm} indicates the $h$ dependence of 
$\sin^2 2\theta_{23}$ as well as $\alpha$. 
The numerical results of solar neutrino mixing angle 
in Fig.~\ref{fig:A1tansol} covers the present experimental 
allowed region rather well. 
\begin{figure}
\begin{center}
\includegraphics[width=5cm,clip]{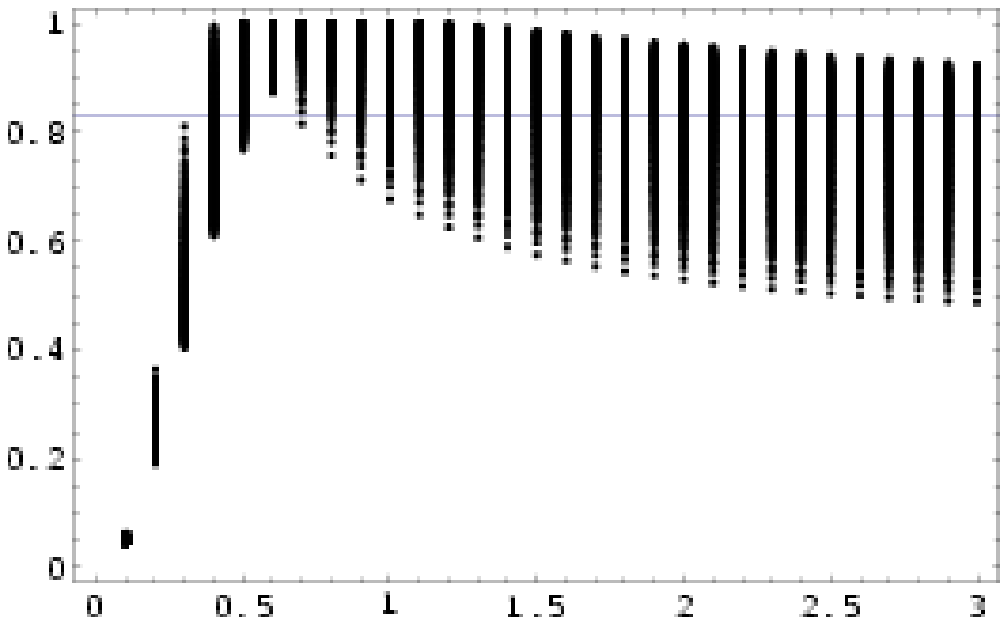}
{\hspace{0.4cm}}
\includegraphics[width=5cm,clip]{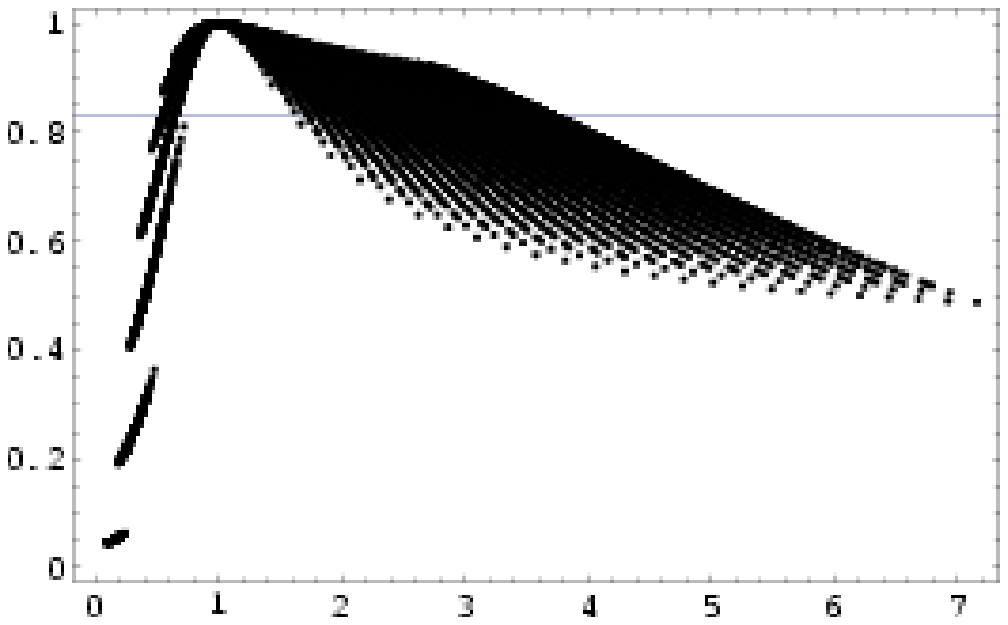}
\end{center}
\caption{Calculated values of $\sin^2 2\theta_{23}$ 
versus $h$ and $\sin^2 2\theta_{23}$ versus $\alpha$ in class $A$.} 
\label{fig:A1sin2atm}
\end{figure}%
\begin{figure}
\begin{center}
\includegraphics[width=5cm,clip]{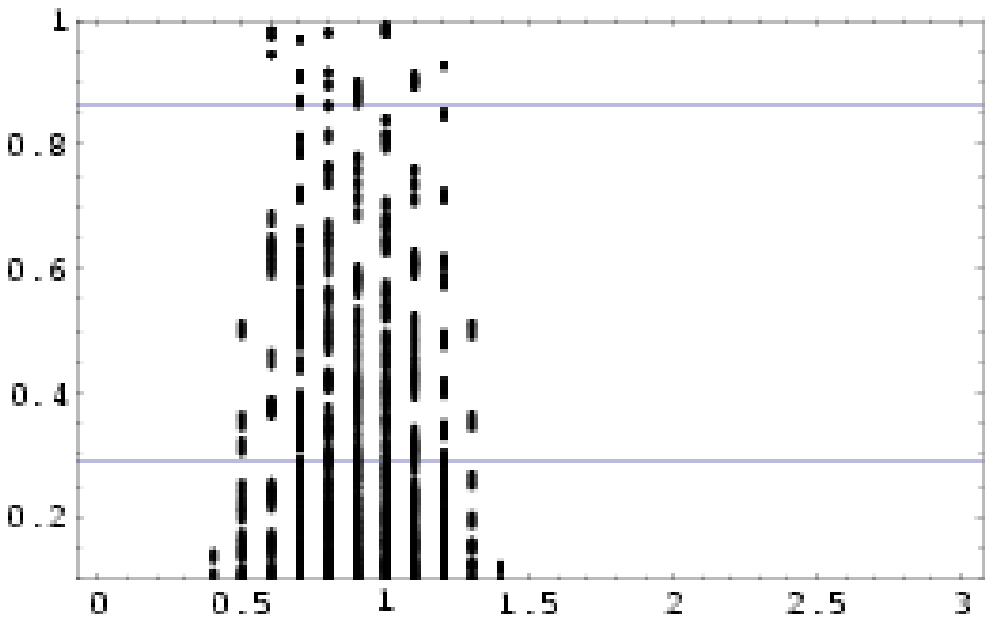}
{\hspace{0.4cm}}
\includegraphics[width=5cm,clip]{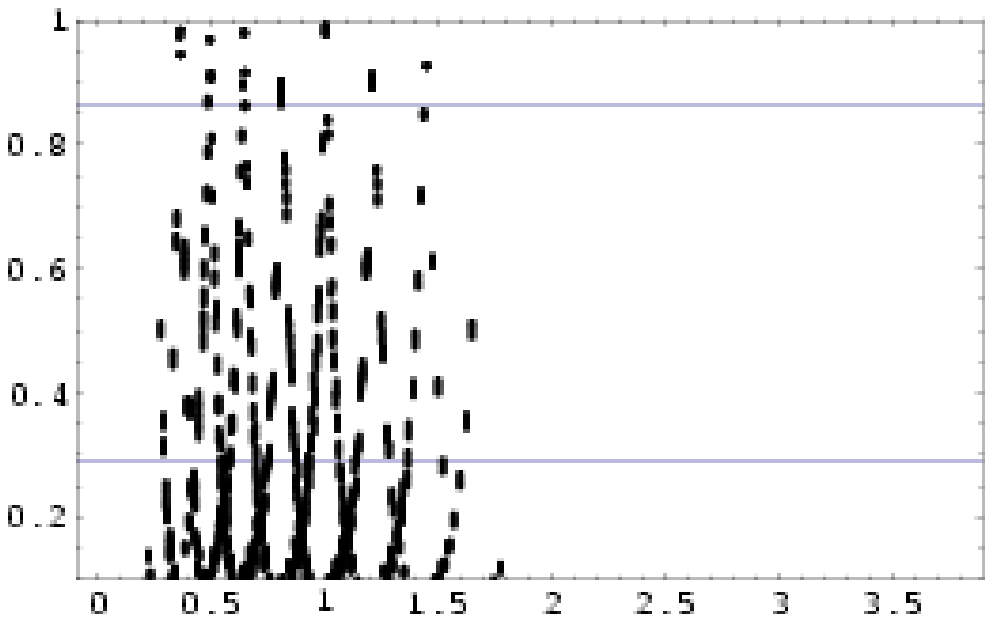}
\end{center}
\caption{Calculated values of $\tan^2 \theta_{12}$ 
versus $h$ and $\tan^2 \theta_{12}$ versus $\alpha$ in class $A$.} 
\label{fig:A1tansol}
\end{figure}%
We further calculate the mass squared differences 
for atmospheric and solar neutrinos, and the ratio of them, 
which are to be compared with the experimental results 
in Eqs. (\ref{expatm}) and (\ref{expsol}). 
These results are showed in Fig.~\ref{fig:A1HMatm} and \ref{fig:A1Hratio}. 
\begin{figure}
\begin{center}
\includegraphics[width=5cm,clip]{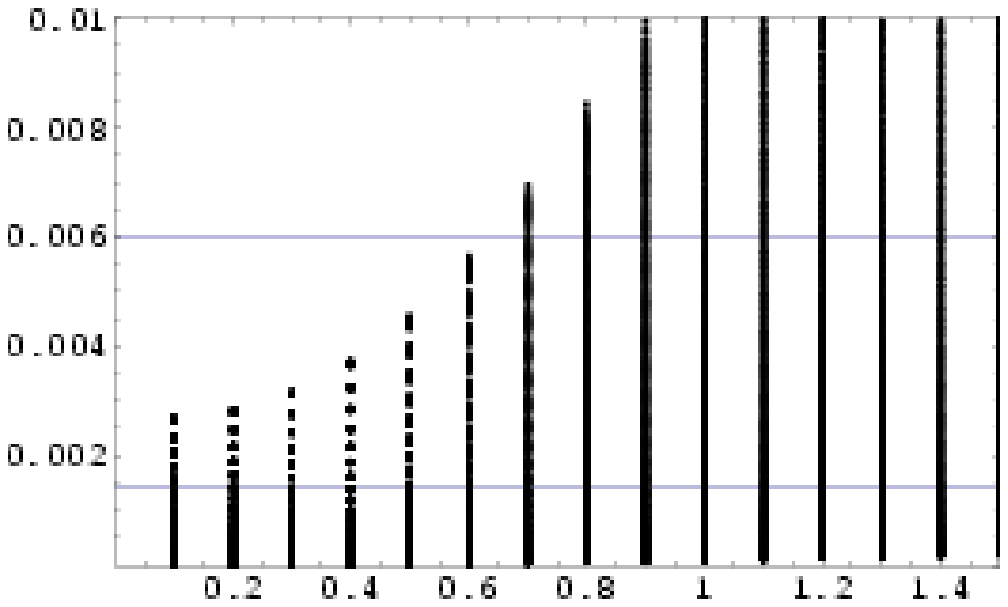}
{\hspace{0.4cm}}
\includegraphics[width=5cm,clip]{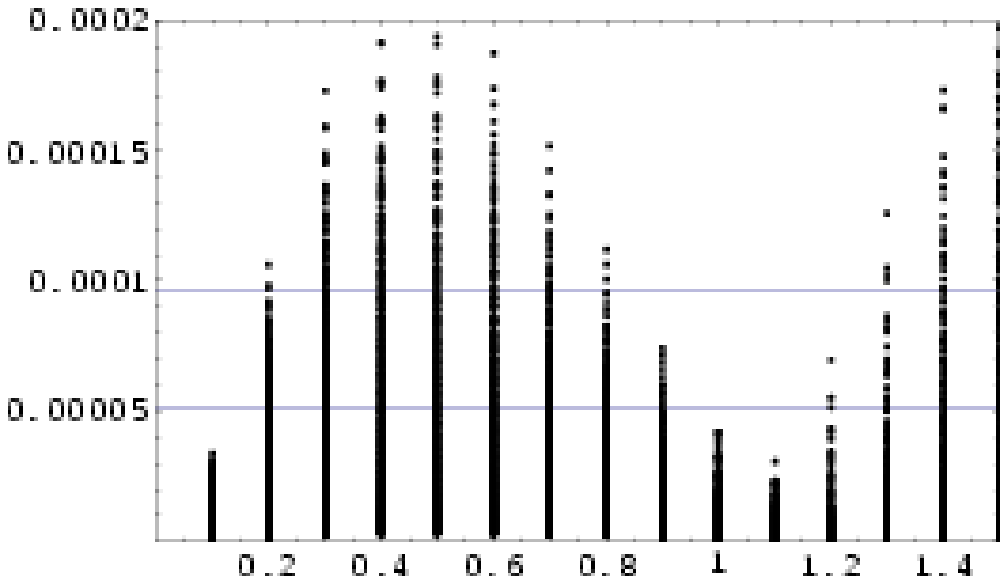}
\end{center}
\caption{Calculated values of $\Delta m^2_{32}$ 
versus $h$ and $\Delta m^2_{21}$ versus $h$ in class $A$.} 
\label{fig:A1HMatm}
\end{figure}%
\begin{figure}
\begin{center}
\includegraphics[width=5cm,clip]{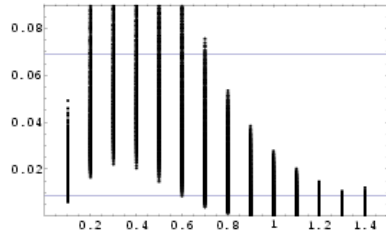}
\end{center}
\caption{Calculated values of $\Delta m^2_{21}/\Delta m^2_{32}$ 
versus $h$ in class $A$.} 
\label{fig:A1Hratio}
\end{figure}%

\subsection{Class $S$}
Finally we have the class $S$ which satisfies both the conditions (i) 
and (ii) as well. 
This class is even easier to reproduce large mixing angle 
$\theta_{12}$ than class $A$,  
since the value of $\beta$ is 
\bea
\beta 
\sim \, -3 \sqrt{\frac{m_c}{m_t}} 
\sim (0.09-0.15)\, h. 
\eea
Fig.~\ref{fig:Ssin2atm} shows the calculated values of $\sin^2 2\theta_{23}$ 
versus $h$ as well as $\alpha$ and 
Fig.~\ref{fig:Stansol} shows the $h$ and $\alpha$ dependence of 
$\tan^2\theta_{12}$. 
From Fig.~\ref{fig:Ssin2atm} and Fig.~\ref{fig:Stansol}, 
one finds the most probable region of $h$ is found within almost $1\pm 0.3$. 
We also show the figure of the calculated values of 
neutrino mass squared differences and their ratio 
(see Fig.~\ref{fig:SHMatm} and Fig.~\ref{fig:SHMratio}). 
All the above results actually indicates that the type $S$ is 
excellently consistent with the present experimental bound. 
\begin{figure}
\begin{center}
\includegraphics[width=5cm,clip]{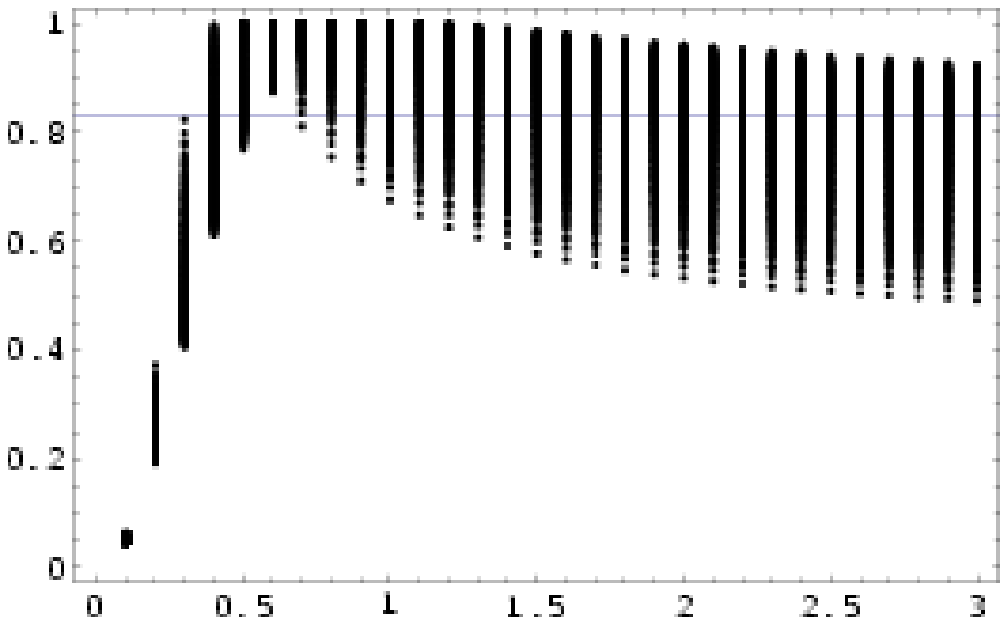}
{\hspace{0.4cm}}
\includegraphics[width=5cm,clip]{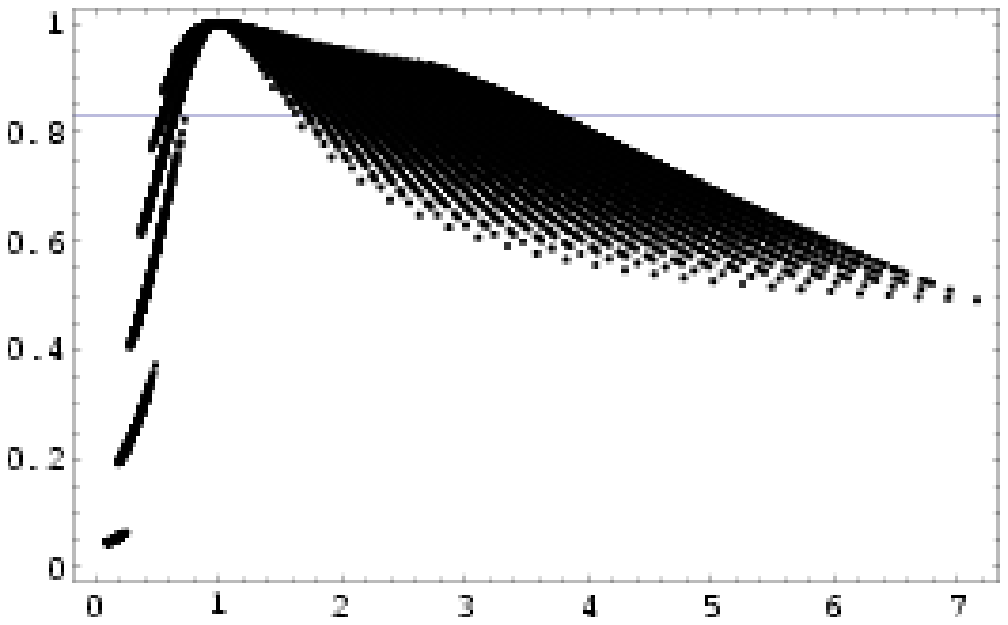}
\end{center}
\caption{Calculated values of $\sin^2 2\theta_{23}$ 
versus $h$ and $\sin^2 2\theta_{23}$ versus $\alpha$ in class $S$.} 
\label{fig:Ssin2atm}
\end{figure}%
\begin{figure}
\begin{center}
\includegraphics[width=5cm,clip]{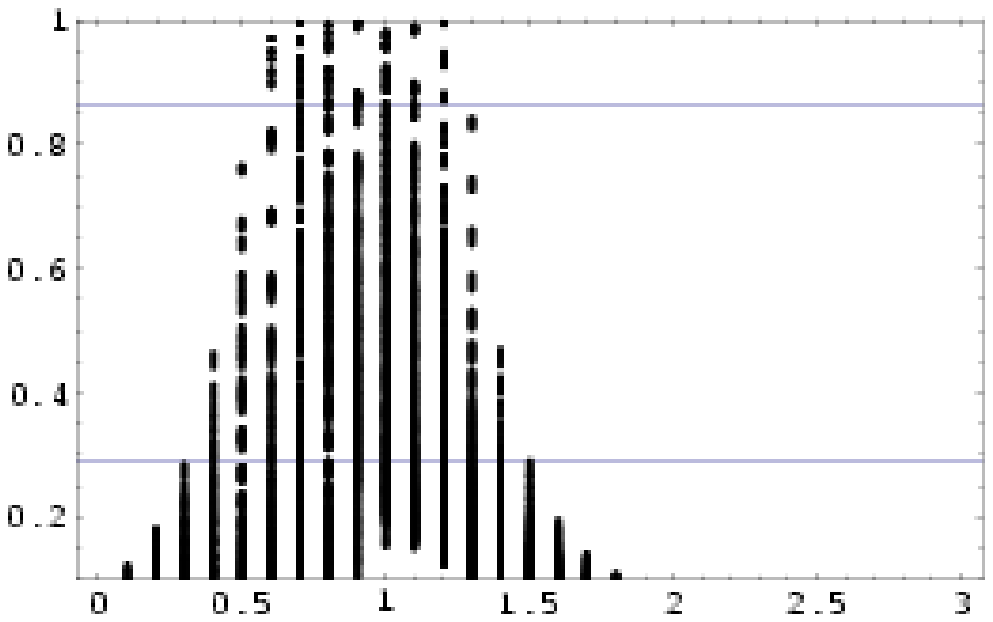}
{\hspace{0.4cm}}
\includegraphics[width=5cm,clip]{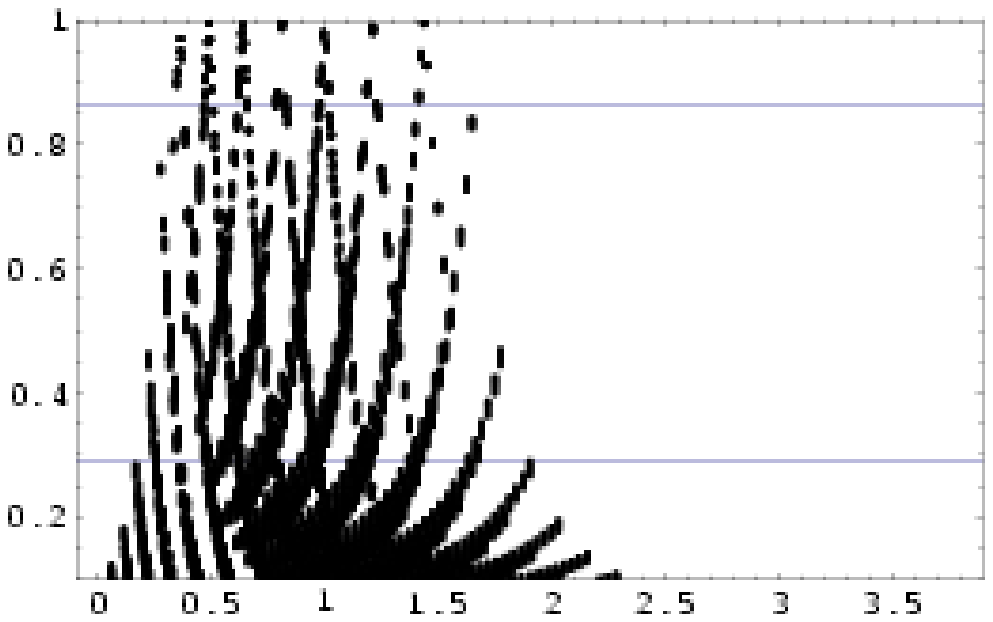}
\end{center}
\caption{Calculated values of $\tan^2 \theta_{12}$ 
versus $h$ and $\tan^2 \theta_{12}$ versus $\alpha$ in class $S$.} 
\label{fig:Stansol}
\end{figure}%
\begin{figure}
\begin{center}
\includegraphics[width=5cm,clip]{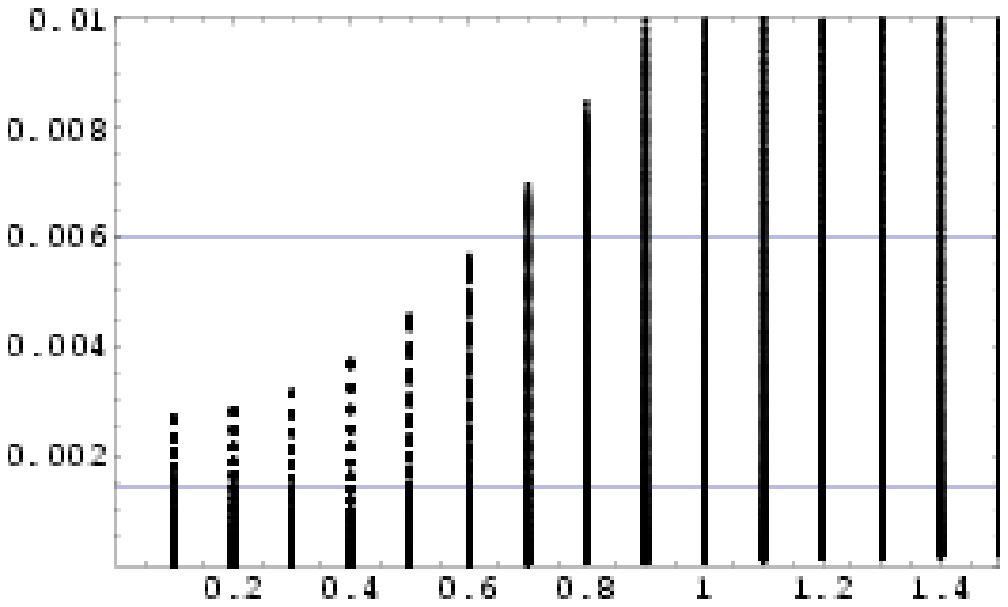}
{\hspace{0.4cm}}
\includegraphics[width=5cm,clip]{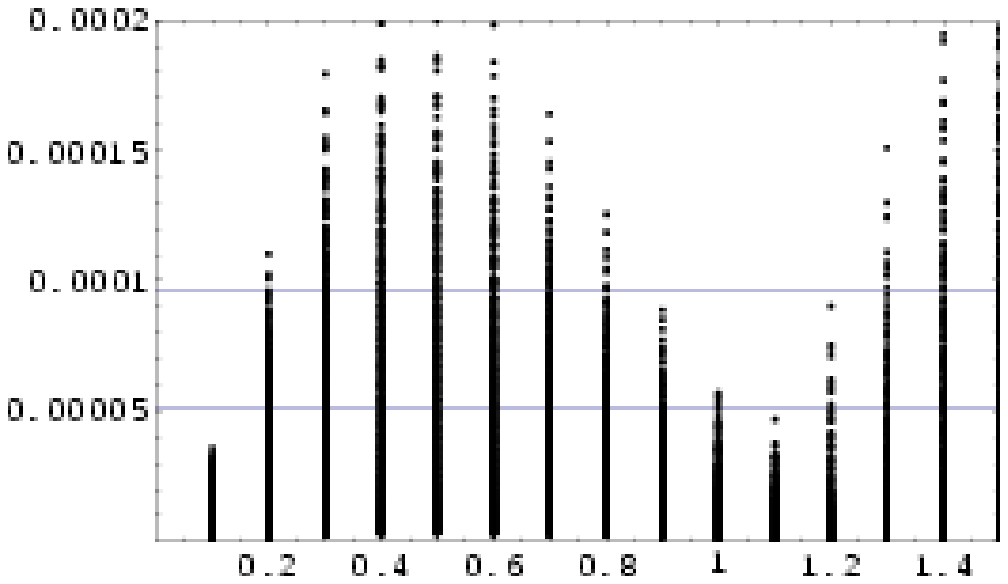}
\end{center}
\caption{Calculated values of $\Delta m^2_{32}$ 
versus $h$ and $\Delta m^2_{21}$ versus $h$ in class $S$.} 
\label{fig:SHMatm}
\end{figure}%
\begin{figure}
\begin{center}
\includegraphics[width=5cm,clip]{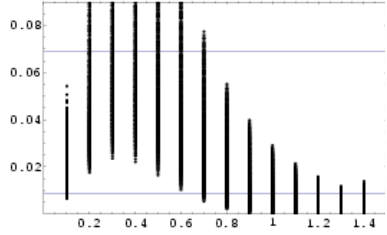}
\end{center}
\caption{Calculated values of $\Delta m^2_{21}/\Delta m^2_{32}$ 
versus $h$ in class $S$.} 
\label{fig:SHMratio}
\end{figure}%

\subsection{Summary}
Here we would like to summarize the range of parameters with which 
we can reproduce the experimental data for the possible candidates, 
the types $S$, $A$ and $B$ in Table~\ref{classificationsa1}, 
and the numerical results for the absolute masses of neutrinos and $|U_{e3}|$ 
in Table~\ref{classificationsa2}.  Also we have summarized the situation of 
the type $C$ and $F$ which are not proper candidates at all 
(see Table~\ref{classificationcf}). 
\begin{table}
\caption{The neutrino properties for the types $S$, $A$ and $B$. 
$\bigcirc$ indicates that corresponding type can reproduce 
the experimental data, 
$\circledcirc$ even better, and $\triangle$ less probable. 
The values of $h$ which can reproduce both $\sin^22\theta_{23}$ and 
$\tan^2 \theta_{12}$ and $m_R$ which can reproduce both $\Delta m^2_{32}$ 
and $\Delta m^2_{21}$ are shown.}
\begin{center}
\setlength{\tabcolsep}{3pt}\footnotesize
\begin{tabular}[b]{c|c|c|c|c|c}\hline

Type & Texture & $\sin^22\theta_{23}$ & $\tan^{2^{\mathstrut}}\theta_{12}$ & 
$h$ & $m_R~(\gev)$ 
\\ \hline 
$S_1$ & 
$\left(
\begin{array}{@{\,}ccc@{\,}}
0 & \textbf{126} & 0 \\
\textbf{126} & \textbf{10} & \textbf{10} \\
0 & \textbf{10} & \textbf{126}
\end{array}
\right)^{\mathstrut}_{\mathstrut}$ &
$\bigcirc$ & $\circledcirc$ & $0.4-1.4$ & $2 \times 10^{15}$ 
\\  \hline

$S_2$ & 
$\left(
\begin{array}{@{\,}ccc@{\,}}
0 & \textbf{126} & 0 \\
\textbf{126} & \ten & \textbf{10} \\
0 & \textbf{10} & \ten 
\end{array}
\right)^{\mathstrut}_{\mathstrut}$ & 
$\bigcirc$ & $\circledcirc$ & $0.4-1.4$ & $2 \times 10^{14}$ 
\\  \hline
$A_1$ & 
$\left(
\begin{array}{@{\,}ccc@{\,}}
0 & \textbf{126} & 0 \\
\textbf{126} & \textbf{126} 
& \textbf{126} \\
0 & \textbf{126} & \textbf{126} 
\end{array}
\right)^{\mathstrut}_{\mathstrut}$ &
$\bigcirc$ & $\bigcirc$ & $0.5-1.3$ & $2 \times 10^{15}$ 
\\  \hline
$A_2$ & 
$\left(
\begin{array}{@{\,}ccc@{\,}}
0 & \ten & 0 \\
\ten & \ten & \ten \\
0 & \ten & \textbf{126} 
\end{array}
\right)^{\mathstrut}_{\mathstrut}$ & 
$\bigcirc$ & $\bigcirc$ & $0.5-1.3$ & $2 \times 10^{15}$ 
\\  \hline
$A_3$ & 
$\left(
\begin{array}{@{\,}ccc@{\,}}
0 & \textbf{126} & 0 \\
\textbf{126} & \textbf{126} & \textbf{126} \\
0 & \textbf{126} & \ten 
\end{array}
\right)^{\mathstrut}_{\mathstrut}$ & 
$\bigcirc$ & $\bigcirc$ & $0.5-1.3$ & $2 \times 10^{14}$ 
\\  \hline
$A_4$ & 
$\left(
\begin{array}{@{\,}ccc@{\,}}
0 & \ten & 0 \\
\ten & \ten & \ten \\
0 & \ten & \ten 
\end{array}
\right)^{\mathstrut}_{\mathstrut}$ & 
$\bigcirc$ & $\bigcirc$ & $0.5-1.3$ & $2 \times 10^{14}$ 
\\  \hline
$B_1$ & 
$\left(
\begin{array}{@{\,}ccc@{\,}}
0 & \ten & 0 \\
\ten & \textbf{126} & \textbf{126} \\
0 & \textbf{126} & \textbf{126} 
\end{array}
\right)^{\mathstrut}_{\mathstrut}$ & 
$\bigcirc$ & $\triangle$ & $0.6-1.2$ & $2 \times 10^{15}$ 
\\  \hline
$B_2$ & 
$\left(

\begin{array}{@{\,}ccc@{\,}}
0 & \ten & 0 \\
\ten & \textbf{126} & \textbf{126} \\
0 & \textbf{126} & \ten 
\end{array}
\right)^{\mathstrut}_{\mathstrut}$ & 
$\bigcirc$ & $\triangle$ & $0.6-1.2$ & $2 \times 10^{14}$ 
\\  \hline
\end{tabular}
\end{center}
\label{classificationsa1}
\end{table}%
\begin{table}
\caption{The predicted values of $|m_{\nu_3}|$, $|m_{\nu_2}|$, 
$|m_{\nu_1}|$ and $|U_{e3}|$ for each type with $h$, $m_R$}
\begin{center}
\setlength{\tabcolsep}{3pt}\footnotesize
\begin{tabular}[t]{c|c|c|c|c|c|c}\hline
Type & $h$ & $m_R~(\gev)$ & 
$~~|m_{\nu_3}|^{\mathstrut}_{\mathstrut}~(\ev)~$ & 
$~~|m_{\nu_2}|~(\ev)~$ & $|m_{\nu_1}|~(\ev)$ & $|U_{e3}|$ \\ \hline 
$S_1$ & $0.4-1.4$ & $2 \times 10^{15^{\mathstrut}}$ &$0.005-0.17$ & 
$0.001-0.015$ & $3 \times 10^{-5}-4 \times 10^{-3}$ & $0.01-0.06$ \\  \hline
$S_2$ & $0.4-1.4$ & $2 \times 10^{14^{\mathstrut}}$ & $0.005-0.17$ & 
$0.001-0.015$ & $3 \times 10^{-5}-4 \times 10^{-3}$ & $0.01-0.06$ \\  \hline
$A_1$ & $0.5-1.3$ & $2 \times 10^{15^{\mathstrut}}$ & $0.006-0.15$ & 
$0.002-0.014$ & $3 \times 10^{-6}-1 \times 10^{-3}$ & $0.006-0.02$ \\  \hline
$A_2$ & $0.5-1.3$ & $2 \times 10^{15^{\mathstrut}}$ & $0.006-0.15$ & 
$0.002-0.014$ & $3 \times 10^{-6}-1 \times 10^{-3}$ & $0.006-0.02$ \\  \hline
$A_3$ & $0.5-1.3$ & $2 \times 10^{14^{\mathstrut}}$ & $0.006-0.15$ & 
$0.002-0.014$ & $3 \times 10^{-6}-1 \times 10^{-3}$ & $0.006-0.02$ \\  \hline
$A_4$ & $0.5-1.3$ & $2 \times 10^{14^{\mathstrut}}$ & $0.006-0.15$ & 
$0.002-0.014$ & $3 \times 10^{-6}-1 \times 10^{-3}$ & $0.006-0.02$ \\  \hline

$B_1$ & $0.6-1.2$ & $2 \times 10^{15^{\mathstrut}}$ & $0.006-0.13$ & 
$0.002-0.014$ & $5 \times 10^{-7}-3 \times 10^{-3}$ & $0.002-0.007$ \\  \hline
$B_2$ & $0.6-1.2$ & $2 \times 10^{14^{\mathstrut}}$ & $0.006-0.13$ & 
$0.002-0.015$ & $6 \times 10^{-7}-4 \times 10^{-3}$ & $0.002-0.007$ \\  \hline
\end{tabular}
\end{center}
\label{classificationsa2}
\end{table}%

\begin{table}
\caption{The neutrino properties for the types $C$ and $F$.  
The notation $\bigcirc$ indicates that corresponding type can 
reproduce the experimental data and $\times$ impossible. }
\begin{center}
\setlength{\tabcolsep}{3pt}\footnotesize
\begin{tabular}{c|c|c|c|c}\hline
Type & Texture & $\sin^22\theta_{23}$ & 
$\tan^{2^{\mathstrut}}\theta_{12}$ & $h$ 
\\ \hline 
$C_1$ & 
$\left(
\begin{array}{@{\,}ccc@{\,}}
0 & \textbf{126} & 0 \\
\textbf{126} & \textbf{10} & \textbf{126} \\
0 & \textbf{126} & \textbf{126} 
\end{array}
\right)^{\mathstrut}_{\mathstrut}$ & 
$\bigcirc$ & $\times$ & none 
\\  \hline
$C_2$ & 
$\left(
\begin{array}{@{\,}ccc@{\,}}
0 & \ten & 0 \\
\ten & \ten & \textbf{126} \\
0 & \textbf{126} & \textbf{126}
\end{array}
\right)^{\mathstrut}_{\mathstrut}$ & 
$\bigcirc$ & $\times$ & none 
\\  \hline

$C_3$ & 
$\left(
\begin{array}{@{\,}ccc@{\,}}
0 & \ten & 0 \\
\ten & \ten & \textbf{126} \\
0 & \textbf{126} & \ten 
\end{array}
\right)^{\mathstrut}_{\mathstrut}$ & 
$\bigcirc$ & $\times$ & none 
\\  \hline
$C_4$ & 
$\left(
\begin{array}{@{\,}ccc@{\,}}
0 & \textbf{126} & 0 \\
\textbf{126} & \textbf{10} & \textbf{126} \\
0 & \textbf{126} & \ten 
\end{array}
\right)^{\mathstrut}_{\mathstrut}$ & 
$\bigcirc$ & $\times$ & none 
\\  \hline
$F_1$ & 
$\left(
\begin{array}{@{\,}ccc@{\,}}
0 & \textbf{126} & 0 \\
\textbf{126} & \textbf{126} 
& \textbf{10} \\
0 & \textbf{10} & \textbf{126} 
\end{array}
\right)^{\mathstrut}_{\mathstrut}$ & 
$\times$ & $\times$ & none 
\\  \hline
$F_2$ & 
$\left(
\begin{array}{@{\,}ccc@{\,}}
0 & \ten & 0 \\
\ten & \textbf{126} & \ten \\
0 & \ten & \textbf{126}
\end{array}
\right)^{\mathstrut}_{\mathstrut}$ & 
$\times$ & $\times$ & none 
\\  \hline
$F_3$ & 
$\left(
\begin{array}{@{\,}ccc@{\,}}
0 & \ten & 0 \\
\ten & \textbf{126} & \ten \\
0 & \ten & \ten 
\end{array}
\right)^{\mathstrut}_{\mathstrut}$ & 
$\times$ & $\times$ & none 
\\  \hline
$F_4$ & 
$\left(
\begin{array}{@{\,}ccc@{\,}}
0 & \textbf{126} & 0 \\
\textbf{126} & \textbf{126} 
& \textbf{10} \\
0 & \textbf{10} & \ten 
\end{array}
\right)^{\mathstrut}_{\mathstrut}$ & 
$\times$ & $\times$ & none 
\\  \hline
\end{tabular}
\end{center}
\label{classificationcf}
\end{table}%

In summarizing we have found that the best type for configuration which is 
consistent with the experiments is $S_1$. We shall discuss  
this type in detail in the next section.
\section{Predictions of Type $S_1$}
\clean
All the above results are summarized in Table~\ref{classificationsa1} 
and Table~\ref{classificationcf}, 
where all the 16 types are classified into 5 classes, 
$F, C, B, A$ and $S$.  
From these tables one sees that the class $S$ reproduces most 
naturally the large mixing angles 
for both atmospheric and solar neutrinos as well as the ratio of 
atmospheric to solar mass squared differences. 
The explicit forms of up-type mass matrix for the class $S$ are 
seen in  Eq.~(\ref{upmasstypes}), as we expected in section 2. 
Those two types yield the same predictions except for the Majorana mass scale. 
The type $S_1$ requires $m_R \sim 2 \times 10^{15}~\gev$ and 
in the type $S_2$, we have $m_R \sim 10^{14}~\gev$, respectively. 
Thus more desirable one may be the type $S_1$ since it predicts more realistic 
bottom-tau ratio at low energy. 
Thus we here discuss the type $S_1$ in detail, 
namely we take, 
\bea
M_U =
\left(
\begin{array}{@{\,}ccc@{\,}}
 0                 &{\bf 126}           & 0   \\
{\bf 126}           &{\bf 10}            &{\bf 10} \\
 0                 &{\bf 10}            & {\bf 126}
\end{array}\right),
\label{126up}
\eea
which determines $M_{\nu_D} $ as follows:
\bea
M_{\nu_D} = m_t 
\left(
\begin{array}{@{\,}ccc@{\,}}
0 & -3a_u & 0 \\ 
-3a_u & b_u & c_u \\ 
0 & c_u & -3
\end{array}
\right).
\label{126dirac}
\end{eqnarray}
With this Eq. (\ref{126dirac}), we obtain the neutrino mass matrix 
from Eq.~(\ref{seesaw}), 
\bea
M_{\nu} = \left(
\begin{array}{@{\,}ccc@{\,}}
 0  & \frac{a_u^2}{r} & 0   \\
\frac{a_u^2}{r} & \frac{-2a_u b_u}{3r} +\frac{ c_u^2}{9}
& -\frac{ c_u}{3} (\frac{a_u}{r}+1) \\
 0  &-\frac{ c_u}{3} (\frac{a_u}{r}+1) & 1
\end{array}
\right) \frac{9m_t^2}{m_R}. 
\label{nuabcr}
\eea
The value $h\sim \mathcal{O}(1)$ is determined to fit the 
experimental large mixing angle $\theta_{23}$, which determines $r$ 
as follows:
\begin{equation}
r = \frac{ac}{d^2h} \simeq  
- \frac{\> 1 \>}{3} \frac{\sqrt{m_u^2m_c}}{\sqrt{m_t^3}}, 
\quad d=-3, \, a=-3a_u, \,  c=c_u.  
\label{eq:mjratio}
\end{equation}
Then the neutrino mass matrix is written as,  
\bea
M_{\nu}
\sim 
\left(
\begin{array}{@{\,}ccc@{\,}}
 0  & -3h\sqrt{\frac{m_c}{m_t}}  & 0   \\
-3h\sqrt{\frac{m_c}{m_t}} & 2h\frac{m_c}{\sqrt{m_um_t}} 
& h \\
 0  &h & 1
\end{array}
\right) \frac{9m_t^2}{m_R}.
\label{app1nuabcr}
\eea
Now that all the neutrino information are determined 
in terms of $m_u,m_c,m_t$ with $h$ or $r$;
\begin{eqnarray}
\tan^2 2\theta_{23} &\simeq& 
\frac{4h^2}{\Bigl( 1 - h \frac{2m_c}{\sqrt{m_um_t}} \Bigr)^2}, \\
\tan^2 2\theta_{12} &\simeq&
\frac{144\, h^2 \, m_c\cos^2\theta_{23}}
{m_t\, \Bigl( 1 - 2h + h \frac{2 m_c}{\sqrt{m_um_t}} \Bigr)^2}, \\   
\sin \theta_{13}  &\simeq& -\frac{6h}{1 + 2h + h \frac{2 m_c}{\sqrt{m_um_t}}} 
\sqrt{\frac{m_c}{m_t}} \sin\theta_{23}\cos \theta_{12}.
\end{eqnarray}
From Fig.~\ref{fig:Ssin2atm} and \ref{fig:Stansol} 
the most probable value of $h$ is almost 1. 
If one adopts $h=1$, then we have
\bea
M_{\nu}
\sim 
\left(
\begin{array}{@{\,}ccc@{\,}}
 0  & -3\sqrt{\frac{m_c}{m_t}}  & 0   \\
-3\sqrt{\frac{m_c}{m_t}} & 2\frac{m_c}{\sqrt{m_um_t}} 
& 1 \\
 0  &1 & 1
\end{array}
\right) \frac{9m_t^2}{m_R}. 
\label{app2nuabcr}
\eea
From this, we obtain the following equations, 
which is what we have investigated in a previous letter~\cite{oharabando}; 
\begin{eqnarray}
\tan^22\theta_{23} &\simeq&  \frac{1}
{\Bigl(1 - \frac{2m_c}{\sqrt{m_um_t}} \Bigr)^2},  \\
\tan^22\theta_{12} &\simeq& \frac{144\, m_c \cos^2\theta_{23}}
{m_t \Bigl( 1-\frac{2m_c}{\sqrt{m_u m_t}} \Bigr)^2}, \\   
\sin \theta_{13} &\simeq& -2 \sqrt{\frac{m_c}{m_t}} 
\sin\theta_{23} \cos\theta_{12}, 
\end{eqnarray}
from which the following equations are derived as
\bea
\tan^22\theta_{12} &\simeq&
\frac{144m_c}{m_t}\tan^22\theta_{23} \cos^2\theta_{23}, \\
\sin^2 \theta_{13}  &\simeq&
\frac{4m_c}{m_t} \sin^2\theta_{23} \cos^2\theta_{12}.
\eea
These relations seem quite desirable to the present experimental indications. 
Interesting enough is that once we know the atmospheric and 
solar neutrino experiments, $|U_{e3}|$ is predicted without 
any ambiguity coming from the up-quark masses at GUT scale;  
\bea
\sin^2 \theta_{13}  \simeq 
\frac{\tan^22\theta_{12}}{36 \tan^22\theta_{23}} 
\tan^2 \theta_{23}\cos^2\theta_{12},
\eea
which is independent of the uncertainty especially coming from the value, 
$m_t$ at GUT scale. 
Next the neutrino masses are given by
\bea
m_{\nu_3} \simeq \lambda_{\nu_3} \frac{m_t^2}{m_R},\quad 
m_{\nu_2} \simeq  \lambda_{\nu_2} \frac{m_t^2}{m_R},\quad 
m_{\nu_1} \simeq \lambda_{\nu_1} \frac{m_t^2}{m_R}, 
\label{neutmasslam}
\eea
where the renormalization factor ($\sim 1/3$) has been 
taken account to estimate the lepton masses at low energy scale. 
Since $\lambda_{\nu_2} \ll \lambda_{\nu_3} \sim \mathcal{O}(1)$, 
this indeed yields $m_R\sim 10^{16}$ GeV, as many people require. 
On the other hand, $m_{\nu_2}$ and $m_{\nu_1}$ may be the same order; 
they differ only by a factor. 

The numerical results of the absolute values of neutrino masses 
are shown in Fig.~\ref{fig:SHmass} and Fig.~\ref{fig:SHmassTau} and 
the predicted values of $|U_{e3}|$ is shown in Fig.~\ref{fig:SHUe3}.  
\begin{figure}
\begin{center}
\includegraphics[width=5cm,clip]{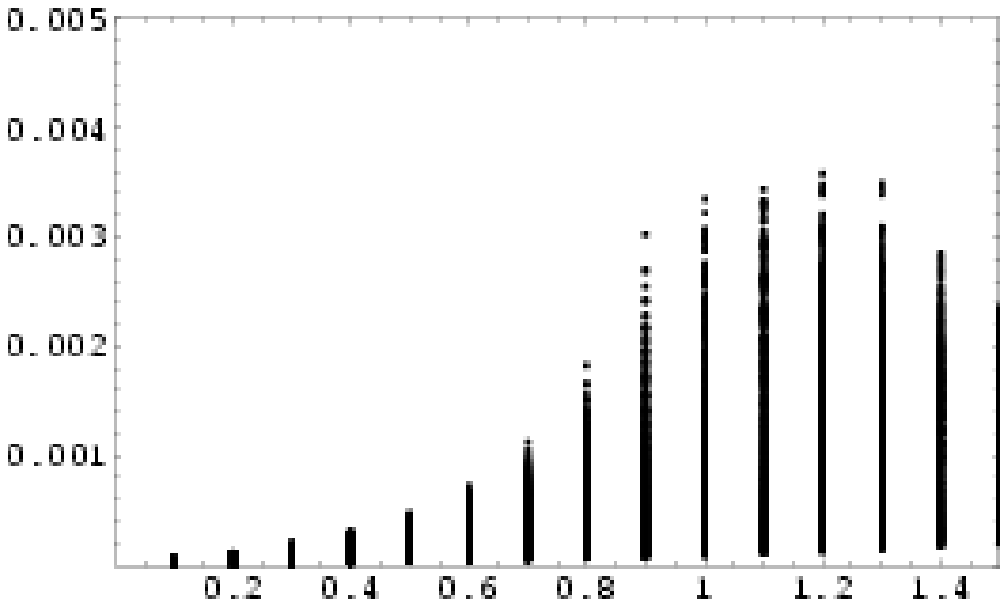}
{\hspace{0.4cm}}
\includegraphics[width=5cm,clip]{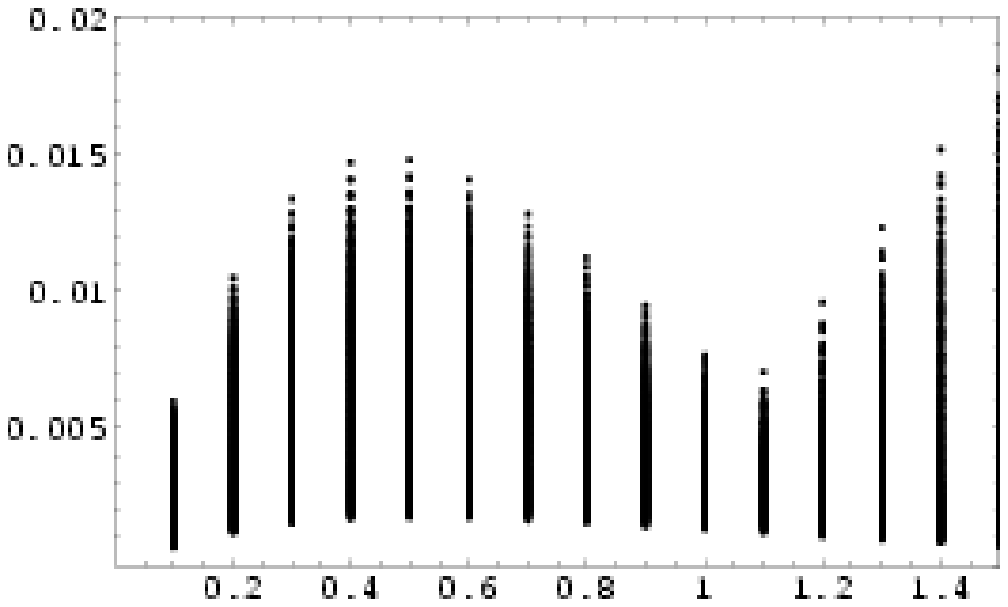}
\end{center}
\caption{Predicted values of $|m_{\nu_1}|$ versus $h$ 
and $|m_{\nu_2}|$ versus $h$ in class $S_1$.} 
\label{fig:SHmass}
\end{figure}%
\begin{figure}
\begin{center}
\includegraphics[width=5cm,clip]{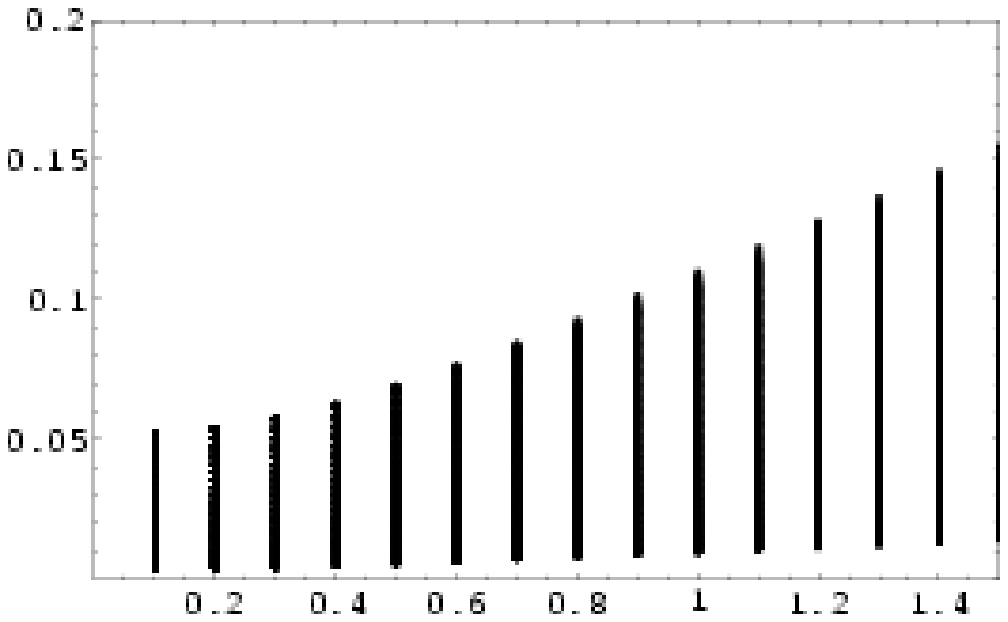}
\end{center}
\caption{Predicted values of $|m_{\nu_3}|$ versus $h$ in class $S_1$.} 
\label{fig:SHmassTau}
\end{figure}%
\begin{figure}
\begin{center}
\includegraphics[width=5cm,clip]{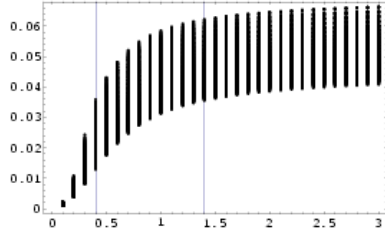}
\end{center}
\caption{Predicted values of $|U_{e3}|$ versus $h$ in class $S_1$.} 
\label{fig:SHUe3}
\end{figure}%
We can predict the values of $|U_{e3}|$,
\bea
|U_{e3}| \sim 0.01-0.06.  
\eea
from Fig.~\ref{fig:SHUe3}.
This value is not so large compared with the contribution
from the charged lepton part, which is of order of 
$\lambda |U_{\mu3}| \sim 0.01$, so the above prediction would
yield additional ambiguity of $\sim 0.01$. 
We hope this can be checked by experiment in near future
JHF-Kamioka long-base line~\cite{JHF}, the sensitivity of 
which is reported as $|U_{e3}| \simeq 0.04$ at $90\%$ C.L.
If we further expect Hyper-Kamiokande ($|U_{e3}|<10^{-2}$)~\cite{Hyper-K}, 
we can completely check whether such symmetric texture model can survive or not. 

In conclusion we list a set of typical values of neutrino masses and 
mixings at $m_t \simeq 240~\gev$: 
\bea
\sin^2 2\theta_{23} &\sim& 0.95-1, \\ 
\tan^2 \theta_{12} &\sim& 0.23-0.6, \\
|U_{e3}|  &\sim& 0.037-0.038, \\
|m_{\nu_3}| &\sim& 0.06-0.07~\rm eV, \\
|m_{\nu_2}| &\sim& 0.003-0.006~\rm eV, \\
|m_{\nu_1}| &\sim& 0.0007-0.0015~\rm eV,
\eea
with $m_R= 2 \times 10^{15}~\gev$ and $r m_R=10^{8}~\gev$, which 
corresponds to the Majorana mass for the third generation and 
those of the second and first generations, respectively.  

Remark that, once the scale of right-handed Majorana mass matrix, $r$,  
is determined so as for the mixing angle of atmospheric neutrinos 
to become maximal, the same value $r$ well reproduces 
the large mixing angle of solar neutrinos and the two mass differences  
$\Delta m^2_{32}$ and $\Delta m^2_{21}$ with $m_R$. 
If we have more exact information of the
experimental values, we need to
include the CP phase factors in order to make more precise
predictions,
which is our next task.
\section{Further Discussions}
\clean

In general there  are four cases as to the origin of 
large mixing for each mixing angle, $\theta_{23}$ and $\theta_{12}$. 
This can be summarized as follows:
\begin{enumerate}
\item[DD]Down Road Option: 
This is first proposed by Yanagida at the Takayama conference 
as a proper realization of non-parallel family structure and may be 
mostly regarded as a natural solution. Many proposals have been made to 
realize such non-parallel family structure~\cite{yanagida,yanagidasato,nomurayanagida}.  
It is indeed successful as to explain a large mixing angle, $\theta_{23}$ naturally. 
In this option we take non-symmetric down-type mass matrix 
under the observation that ${\bf 5^*}$ multiplets in $SU(5)$ GUT 
contains $SU(2)_L$ singlet down-quarks and doublet charged leptons. 
An example is the following  non-symmetric mass matrix~\cite{Habamurayama}  
\begin{eqnarray}
M_l = M_d^{\rm T} \sim
\left(
\begin{array}{@{\,}ccc@{\,}}
\epsilon^2   &   \epsilon^2   &     \epsilon^2  \\
\epsilon     &  \epsilon      &     \epsilon    \\
1     &    1      &    1  
\end{array}
\right).  
\label{danarchy}
\end{eqnarray}
With a small number $\epsilon$ we can provide  a large lepton mixing 
on the one hand, and a small down-quark mixing on the other hand. 
The remarkable fact implies that the mass matrix in the down-type sector is 
not symmetric even in such a large unification group as $E_6$ and that it
leads to quite different lepton and down quark mixing angles.
This is really realized even if we take 
such GUT larger than $E_6$, for example, 
by using E-twisted family structure~\cite{Bando:1999km}. 
\item[DU] Down-Up Option: 
However even if we can naturally reproduce large mixing angle of 
$\theta_{23}$, we need further tuning to get large mixing 
angle for $\theta_{12}$. This is really serious because 
most of the calculations are done only  within 
``order of magnitude'' arguments. Because of this, 
it is very difficult to get exact numbers of small 
parameters of the first and second generations. 
So there might be another possible option in which 
charged lepton mass matrix reproduces  only large 2-3 mixing angle, 
leaving the neutrino mass matrix being 
responsible to derive solar large mixing angle. 
\item[UD] Up-Down  Option: 
This is the option in which $\sin^22\theta_{23}$ is due to 
neutrino mass matrix and large $\tan^2\theta_{12}$ comes from 
charged lepton sector. However if it is so we can write
\begin{eqnarray}
V_{MNS} &=& U_l^{\dagger}U_{\nu} \simeq
\left(
\begin{array}{@{\,}ccc@{\,}}
   \cos \theta_{12}   &  \sin \theta_{12}  &  0  \\
   - \sin \theta_{12} &  \cos \theta_{12}  &  0  \\
            0            &    0                  &  1 
\end{array}
\right)
\left(
\begin{array}{@{\,}ccc@{\,}}
   1     &    0      &   0     \\
   0     &    \cos \theta_{23}  &   - \sin \theta_{23} \\
   0     &    \sin \theta_{23}  &   \cos \theta_{23} 
\end{array}
\right) 
\nonumber  \\
&=& \left(
\begin{array}{@{\,}ccc@{\,}}
    \cos \theta_{12} & \sin \theta_{12} \cos \theta_{23}
                        & -\sin \theta_{12} \sin \theta_{23} \\
  - \sin \theta_{12} & \cos \theta_{12} \cos \theta_{23} 
                        & - \cos \theta_{12} \sin \theta_{23} \\
      0     &  \sin \theta_{23}  &  \cos \theta_{23} 
\end{array}
\right) 
\label{mnsud}
\end{eqnarray}
which automatically induces large $|U_{e3}|$. This is already excluded by 
the CHOOZ experiment~\cite{CHOOZ}: $|U_{e3}|<0.2$. 
\item[UU] Up-Road Option: 
Both large mixing angles may come from neutrino sector. 
Even within $SU(5)$, the up-quark mass matrix is expressed 
in terms of ${\bf 10} \times {\bf 10}$, and simple symmetric texture 
is usually adopted. 
Then the Dirac neutrino mass matrix is also hierarchical with  
small mixing angles. 
A simple texture has been proposed for the right-handed Majorana neutrino 
mass matrix to give a large $\nu_{\tau}$-$\nu_{\mu}$ mixing~\cite{Bando:1997ns}. 
Thus it is quite nontrivial question whether we can 
reproduce two large mixing angles 
if one concentrates on most economical symmetric texture. 
\end{enumerate}  
We have investigated and checked all the types of textures for 
neutrino mass matrix, and fully investigated to confirm that 
the type $S_1$ is most proper one. 
As for the down-type Yukawa couplings we can 
compare the down quark with charged lepton masses and 
a good type has long been well known named Georgi-Jarlskog 
type mass matrix.  On the other hand, until 
neutrino oscillation data were reported, we have had no 
information as to which representation of Higgs fields 
are relevant to up-type Yukawa couplings. 
Now that a good option for the Higgs configuration with 
symmetric 4-zero texture has been found, 
our next task would be to seek for the origin of the texture, 
which may be related to further higher symmetry or 
to the spatial structure including extra dimensions.  
We saw that the texture zero structure is very important 
to reproduce large mixing angles of neutrino mass matrix 
out of very small quark mass matrix. 
Such zero texture may be a reflection of family symmetry. 

Note that our scenario is quite different from the 
down road option 
in which two large mixing angles observed in neutrino 
oscillation data come from the charged lepton side. 
One can predict some relations of neutrino mixing angles in terms of 
down quark information~\cite{Bando:2000at}, for example.  
However even in such situation we can no more predict
the absolute values of neutrino masses, which indeed 
needs the information of $M_{\nu}$. 
In the up road option, on the other hand, a remarkable fact is 
that we can predict all the neutrino (absolute) masses 
and mixing angles without any ambiguity. 
Our scenario, if it is indeed true, can be checked 
without any ambiguity even for the order-one coefficients. 
The remarkable results are obtained really thanks to the power of GUT.
Our remaining task is to investigate the full 
mass matrix including CP phase. The GUT relation may simplify our analysis. 

\section*{Acknowledgements}
This work started from the discussion 
at the research meeting held in Nov. 
2002 supported by the Grant-in Aid for Scientific Research
No. 09640375. 
We would like to thank to M.~Tanimoto,  A.~Sugamoto and T.~Kugo 
whose stimulating discussion encouraged us very much. 
Also we are stimulated by  the fruitful and instructive discussions 
during the Summer Institute 2002 held at 
Fuji-Yoshida. 
M.~B.\  is supported in part by
the Grant-in-Aid for Scientific Research 
No.~12640295 from Japan Society for the Promotion of Science, and 
Grants-in-Aid for Scientific Purposes (A)  
``Neutrinos" (Y.~Suzuki) No.~12047225, 
from the Ministry of Education, Science, Sports and Culture, Japan.
%
%
%
%

\end{document}